\documentclass[referee]{aa} 

\usepackage{graphicx}
\usepackage{txfonts}
\usepackage{natbib}
\bibpunct{(}{)}{;}{a}{}{,}

\begin{document}

\newcommand{\pd}[2]{\frac{\partial #1}{\partial #2}}
\newcommand{\A}[1]{\boldsymbol{#1}}

\title{Formation of dynamical structures in relativistic jets: the FRI case}

\author{P. Rossi \inst{1}
\and A. Mignone\inst{1,2} 
\and G. Bodo\inst{1} 
\and S. Massaglia\inst{2} 
\and A. Ferrari\inst{2,3} 
}

\offprints{P. Rossi}

\institute{INAF/Osservatorio Astronomico di Torino, Strada Osservatorio 20,
           10025 Pino Torinese, Italy
\and
Dipartimento di Fisica Generale, Universit\`a degli Studi di Torino, via P. Giuria 1, 
10125 Torino, Italy
\and
Department of Astronomy and Astrophysics, University of Chicago, 5640 S. Ellis, Chicago, IL 60637}

\date{Received ; accepted }

\abstract
{Strong observational evidence indicates that all extragalactic jets 
 associated with AGNs move at relativistic speed up to $100$ pc - 1 kpc 
 scales from the nucleus. At larger distances, reflecting the 
 Fanaroff-Riley radio source classification, we observe an abrupt
 deceleration in FR-I jets while relativistic motions persist up to Mpc 
 scale in FR-II. Moreover, VLBI observations of some object like B2 1144+35, 
 Mrk501 and M87 show limb brightening of the jet radio emission at the parsec scale.
 This effect is interpreted kinematically as due to the presence of a deboosted
 central spine at high Lorentz factor and of a weakly relativistic external layer. }
{In this paper we investigate whether these effects can be interpreted
 by a breaking of the collimated flow by external medium entrainment
 favored by shear instabilities, namely Kelvin-Helmholtz instabilities. 
 We examine in details the physical conditions under which significant deceleration 
 of a relativistic flow is produced.}
{We investigate the phenomenon by means of high-resolution three-dimensional 
 relativistic hydrodynamic simulations using the PLUTO code for 
 computational astrophysics.}
{We find that the parameter of utmost importance in determining the instability evolution
 and the entrainment properties is the ambient/jet density contrast. We show that lighter
 jets suffer stronger slowing down in the external layer than in the central part and conserve a 
 central spine at high Lorentz factor.}
{Our model is verified by constructing synthetic emission maps from the numerical simulations that 
 compare reasonably well with VLBI observations of the inner part of FR-I
 sources.}


\authorrunning{Rossi et al.}
\titlerunning{Dynamical structures in relativistic jets}
\maketitle

\section{Introduction}
%
%
%
%
%
%

Extragalactic radio sources are traditionally divided in two morphological classes
according to their intrinsic power \citep{FR74}: low luminosity sources
(Fanaroff-Riley type I, FR-I) are brighter close to the nucleus of the
parent galaxy and their jets become dimmer with distance, while high power sources 
(Fanaroff-Riley type II, FR-II) show the maximum brightness in the hot spots at the jet
termination. The different morphology is generally accepted to reflect a difference in how 
the jet energy is dissipated during propagation in the extragalactic medium and produces the observed
radiation. For FR-I sources, it was quickly accepted that
entrainment and deceleration of the jet must play an important role 
in shaping their morphology \citep{Bicknell84, Bicknell86, DeYoung96,
Komissarov94}, while in FR-II sources energy and momentum are  
transported without losses to the front working surface. 

More recently a large body of evidence has accumulated
showing that jets are relativistic at their base, not only in FR-II
radio sources but also in FR-I. 
Superluminal motions are observed on
milliarcsecond scales in several FR-I jets \citep{Giovannini01} and on
arcsecond scales in M87 and Cen A \citep{Biretta95, Hardcastle03} and
finally, on small scales, there are also observations of one-sidedness
and brightness asymmetry between jet and counter-jet, due most likely
to Doppler boosting effects \citep{Laing99}. FR-I sources are also thought to be the
parent population of BL Lac objects, for which the presence of
relativistic velocities on parsec scales is well established
\citep{UP95}.  The Lorentz factors of
the jet bulk motion at sub-pc scales cannot be deduced directly from the
observations, but are inferred from assumption   
on the physical emission mechanism. \citet{Harris06} in their review
indicate that FR-II radiogalaxies have jets with bulk Lorentz factors
between 5 and 40 \citep[see also][]{UP95} and that FR-I jets have less
constrained values somewhat lower than those of the FR-II jets. \citet{Giovannini01}
 using data on
proper motions and  brightness ratio between jet and counter jet in a
sample of radiogalaxies  do not  find any  systematic difference between
low and high power radio sources, and the values they derive are
between 3 and 10. \citet{CG08} derive  jet
physical properties modelling the spectral energy distribution in a
sample of blazars and  find no difference between BLLac's
(associated to low power radio sources in the unified model) and
radio-quasars (associated to high power radio sources), with an
average value of about 15. 

On the other hand, relativistic motions from the inner regions all the way to larger scales, 
with Lorentz factors of about ten, appear  to be present in powerful
FR-II jets, as indicated by Chandra discovery of bright X-ray emission
at kpc scales \citep{Tavecchio04, Harris06}; instead sub-relativistic velocities are found at kpc scales
\citep{Bicknell94} in low power radio-sources, and the decrease of brightness
asymmetry along the jet suggests that a deceleration must occur \citep{Laing99, Laing02, Laing05}. 

Both these morphological and kinematical data are indicative of the interaction between 
collimated outflows and the surrounding medium. In particular
jet deceleration can be obtained by redistributing the bulk momentum 
through some form of mass entrainment  \citep{Bicknell94, Bicknell95}.  
Velocity shear instabilities are the most likely triggering mechanisms of entrainment
\citep{DeYoung96, DeYoung05}. \footnote{An alternative and most likely
complementary view for the origin of the entrainment has been
presented by \citet{Komissarov94} and \citet{Komissarov96}, who consider the
possibility of entrainment by injection of mass lost by stars within
the jet volume. } In fact the nonlinear development of velocity shear or
Kelvin-Helmholtz instabilities leads to an exchange of mass, momentum
and energy at contact discontinuities between fluids in relative motion \citep{khcyl94,
slab, jet3d}; understanding the details of this process is
essential for modeling the jet dynamics and for giving 
clues on the determination of the jet physical parameters. One of the
key features of this form of interaction is the formation of a mixing
layer at the interface between jet and surrounding medium, where the
external material is entrained and accelerated at the expenses of the
jet momentum. This process leads to the formation of a transverse
velocity profile where the internal layers feel less the effects of
the interaction with the ambient medium and keep an higher velocity,
while the external layers are more decelerated. The presence of such
transverse velocity structure has been already suggested for
explaining some of the observational properties of radio sources such
as their magnetic field configuration \citep{Komissarov90, Laing93},
limb brightening effects \citep{Giroletti04} and to overcome problems
in unifying radiogalaxies with BL Lac objects \citep{Chiaberge00}. 

For studying in detail the instability evolution and the subsequent
entrainment process one has to resort to three-dimensional numerical
simulations, since the mechanisms at the base of the entrainment are
inherently three-dimensional, as shown in preliminary
analyses by \cite{BodoNAR03} and \cite{Rossi04}. 

In this paper we present results of high-resolution hydrodynamic simulations  
in which we follow the evolution of a perturbed 
relativistic jet as it propagates in a homogeneus stationary ambient medium. 
We assume that the jet is in pressure equilibrium with the outside medium, 
although this choice is not critical for the final results.
Perturbations
grow as a consequence of the velocity shear instabilities and lead to
entrainment of external medium and to jet deceleration. The main
questions we address are the dependence of the
entrainment and deceleration processes on the jet physical
parameters and the kind of structure that the jet acquires as the
result of these processes.

The plan of the paper is the following: in \S\ref{sec:numsetup} we 
present the numerical setup adopted for calculations and the parameter space
covered by the simulations; in \S\ref{sec:results} we discuss the results of our 
simulations focusing our attention on the dependence of the efficiency of deceleration
on the physical parameters; in \S\ref{sec:entr} we
analyze in more detail the entrainment properties for the case that appears 
more successful in decelerating
the jet. Preliminary comparisons of our simulations with FR-I source 
of different morphologies are reported in \S\ref{sec:implic}
and conclusions of our study are drawn in \S\ref{sec:disc}.

\section{Numerical setup}
\label{sec:numsetup}
%
%
%
%
%
%

Numerical simulations are carried out by solving 
the equations of particle number and energy-momentum conservation.
Referring to the observer's reference frame, where the fluid moves with 
velocity $v_k$ (in units of the speed of light $c$) with respect to the coordinate 
axis $k=x,y,z$ and assuming a flat metric, the 
conservation laws take the differential form:  
\begin{equation}\label{eq:conslaw}
 \pd{}{t}\left(\begin{array}{c}
  \rho\gamma \\ \noalign{\medskip}
  w \gamma^2 v_k  \\ \noalign{\medskip}
  w \gamma^2 - p  \\ \noalign{\medskip} 
  \rho\gamma f  
\end{array}\right) 
  +
 \sum_i\pd{}{x_i} \left(\begin{array}{c}
 \rho\gamma v_i                   \\ \noalign{\medskip}
 w\gamma^2 v_k v_i + p\delta_{ki} \\ \noalign{\medskip}
 w \gamma^2 v_i                   \\ \noalign{\medskip}
 \rho\gamma f v_i
\end{array}\right) = 0 \,,
\end{equation}
where $\rho, w, p$ and $\gamma$ denote, respectively, the rest 
mass density, enthalpy, gas pressure and Lorentz factor.
The jet and external material are distinguished using a passive tracer, $f$, 
set equal to unity for the injected jet material and equal to zero for the 
ambient medium. The system of equations (\ref{eq:conslaw}) 
is completed by specifying an equation of state relating $w$, $\rho$ and $p$. 
Following \cite{Mignone_et_al05}, we adopt the following prescription: 
\begin{equation}
 w = \frac{5}{2}p + \sqrt{\frac{9}{4}p^2 + \rho^2}
\end{equation}
which closely reproduces the thermodynamics of the Synge gas for a 
single-specie relativistic perfect fluid offering, at the same time, 
considerable ease of implementation and reduced numerical cost
\citep{Mignone_McKinney07}.

Simulations are carried out on a Cartesian domain with coordinates 
in the range $x\in [-L_x/2,L_x/2]$, $y\in [0,L_y]$ and $z\in [-L_z/2,L_z/2]$
(lengths are expressed in units of the jet radius, $y$ is the direction of jet propagation).
At $t=0$, the domain is filled with a perfect fluid of uniform density 
and pressure values representing the external ambient medium initially at rest.
A constant cylindrical inflow is initialized in the small region 
$y \le 1,\,\sqrt{x^2 + z^2} \le 1$ with velocity along the $y$ coordinate 
and is constantly fed into the domain through the lower $y$ boundary.
Jet inflows are represented by a three-parameter family distinguished by the
beam Lorentz factor $\gamma_b$, Mach number $M=v_b/c_s$ and the 
ambient-to-beam density contrast $\eta$ as seen in the observer's
frame. 
Thus, by normalizing the beam density to $\rho_b\gamma_b=1$, the ambient 
medium is given by $\rho_a = \eta/\gamma_b$. The beam is initially 
pressure-matched and $p_b$ is recovered from the definition of the
Mach number.  This assumption is the most used in jet simulations 
  and in this first study we maintain it, 
  however it can be questiond, since it is not clear what could be 
   the mechanism that brings pressure
  equilibrium between the jet and the surrounding medium.
  An alternative approach is suggested by \citet{KomissarovFalle98}
  and could be worth investigating in future studies. 
  In our simulations we fixed the jet Lorentz factor
  equal to 10 at the jet inlet, a representative value in the range
  derived from the observations discussed in the Introduction. For
  the Mach number we choose the values of 3 and 30; however for relativistic
  fluids it is actually more appropriate to consider the relativistic
  Mach number \citep{Konigl80, KomissarovFalle98} defined as $M_{r} = \gamma_{b} v_{b}
  / \gamma_{s} c_{s}$, where $\gamma_{s} =
  1/\sqrt{1-c_{s}^{2}/c_{2}}$. It is in fact $M_{r}$ that has the same
  physical interpretation as the usual non-relativistic Mach number since it reflects the
  propagation of sound waves and the value of Mach angle. Finally for the
  density ratio we choose the values $10^{2}$ and $10^{4}$. The complete
set of parameters for our simulations is given in Table 1.


At the jet inlet, transverse velocities are perturbed introducing pinching,
helical and fluting modes with corresponding azimuthal ($\phi= \tan^{-1}(z/x)$) 
wave numbers $m=0,1,2$:
\begin{equation}
 \left(v_x,v_z\right) = 
\frac{A}{24} \sum_{m=0}^{2} \sum_{l=1}^{8}  
      \cos\left(m\phi + \omega_lt + b_l\right)
 \left(\cos\phi,\sin\phi\right),
\end{equation}
where high ($l=1,\dots, 4$) and low ($l=5,\dots, 8$)
frequency modes are given, respectively, by $\omega_l = c_s(1/2, 1, 2, 3)$ and 
$\omega_l = c_s(0.03, 0.06, 0.12, 0.25)$.
The phase shifts $b_l$ are randomly chosen.
The amplitude $A$ of the perturbation corresponds to a fractional
change in the bulk Lorentz factor, $\gamma_b\to\gamma_b(1+\epsilon)$, 
yielding  
\begin{equation}
 A =  \frac{\sqrt{(1+\epsilon)^2 - 1}}{\gamma_b(1+\epsilon)}\,.
\end{equation}
We typically set $\epsilon = 0.05$.
Outside the inlet region we impose symmetric boundary conditions (emulating the presence of a counter-jet), 
whereas the flow can freely leave the domain throughout the remaining boundaries. 

Explicit numerical integration of the equations (\ref{eq:conslaw}) is achieved
using the relativistic hydrodynamics module available in the 
PLUTO code \citep{PLUTO}. 
PLUTO is a Godunov-type, multi-physics code providing a variety of computational algorithms 
for the numerical integration of hyperbolic conservation laws in multiple spatial 
dimensions. 
For the present application, we employ the dimensionally split version of the 
relativistic Piecewise Parabolic Method (PPM) presented in \cite{Mignone_et_al05},
which has an overall $2^{\rm nd}$ order accuracy in both space and time. 

The physical domain is covered by $N_x\times N_y\times N_z$ computational zones,
not necessarily uniformly spaced.
For domains with a large physical size, we employ a uniform grid resolution around 
the beam (typically for $|x|,|z| \le 2$) and a geometrically stretched grid elsewhere.

\begin{table*}  
\centering
\begin{tabular}{crrrrccc}\hline                                        
Case & $\gamma_b$ & M    & $M_r$ & $\eta$        & pts/beam & $L_x\times L_y\times L_z$   & $N_x\times N_y\times N_z$ \\
\hline\hline\noalign{\medskip}
A    & $10$     & $3$  & 28.3 & $10^2$        & $20$     & $50 \times 150\times 50$  & $324\times1200\times324$ \\
B    & $10$     & $3$  & 28.3 & $10^4$        & $20$     & $60 \times  75\times 60$  & $344\times600\times344$  \\
C    & $10$     & $3$  & 28.3 & $10^4$        & $12$     & $50 \times  75\times 50$  & $172\times300\times172$  \\   
D    & $10$     & $30$ & 300 & $10^4$        & $20$     & $50 \times 150\times 50$  & $324\times1050\times324$ \\
E    & $10$     & $30$ & 300 & $10^2$        & $12$     & $24 \times 200\times 24$  & $144\times560\times144$  \\
\noalign{\medskip}
\hline
\end{tabular}
\caption[]{Parameter set used in the numerical simulation model, the second 
column refers to Lorentz factor $\gamma_b$, the third to the Mach number, the 
fourth to the relativistic Mach number, the fith to the ratio of
proper densities, 
the sixth to number of mesh points on jet radius, 
the seventh the physical domain in units of jet radii and the last one the numerical 
domain}
\label{parset}
\end{table*}

\section{Results}
\label{sec:results}
%
%
%
%
%
%

The overall characteristics of relativistic jet propagation have been originally 
discussed  by \citet{Marti97} who covered a wide
portion of the parameter space with 2D simulations. 
Three-dimensional results for one of the cases studied by
\citet{Marti97} have been presented by \citet{Aloy99}.
Their simulations generalize to the relativistic case typical results of the non-relativistic regime, 
namely: (i) lighter jets exhibit fatter, 
nearly spherical cocoons (see e.g. Krause \& Camenzind 2003) as a 
consequence of the reduced head propagation velocity and 
bulk momentum of the beam; (ii) heavier jets advance
at higher velocities generating cylindrical, thinner cocoons; (iii) high Mach
numbers have the same effect as low density contrasts in producing
elongated  (spearhead) cocoons, while low Mach numbers contribute to form
fatter cocoons \citep{Massaglia96}.  \citet{Marti97} show that a similar behavior is found
for relativistic jets, the main difference being that, increasing the
Lorentz factor, jets, even at low density, tend to form  cocoons
that resemble those of the heavy non-relativistic case due to their
increased inertia. In addition in relativistic fluid dynamics,
different Mach numbers correspond to different jet  
internal energies rather than beam velocities, thus being 
the density contrast  $\eta$ the main parameter defining the shape of the cocoon.

Our results, as we shall see, confirm this global trend; however,
in this scenario, we intend to focus our attention on the
entrainment properties as a possible source of the jet deceleration.
The entrainment process takes place from the interaction between the 
jet beam and the cocoon, promoted by the development of
Kelvin-Helmholtz instabilities (KHI henceforth) at the beam interface. 
The entrained material is composed by jet backflowing material mixed with 
the shocked ambient medium through the contact discontinuity. 
The process is therefore quite complex and determined by many different
factors, the behavior of KHI being of course one of the most relevant.  
The linear analysis of the KHI for relativistic flows 
\citep{Hardee87, Hardee98, Hardee00} shows that the growth rate 
of unstable modes depends on the Mach number, 
density ratio between jet and external medium and Lorentz factor. 
In this respect, we intend to investigate for which set of parameters 
entrainment occurs more efficiently.

Obviously the numerical resolution is a very important factor in simulating the mixing processes. 
Therefore we perform our study with very high-resolution 3D computations in the 
hydrodynamic regime. The presence of magnetic field may change the global structure, 
but as long as it is not too strong it is not believed to change the diffusion processes. 
Given the high computational cost
of 3D simulations, we had to limit our study to a fixed value of the Lorentz
factor varying only the Mach number and density ratio.
Table 1 lists the complete set of parameters adopted in our numerical 
simulations, together with the mesh resolution.
\begin{figure*} 
\begin{tabular}{l}
\includegraphics[scale=0.5]{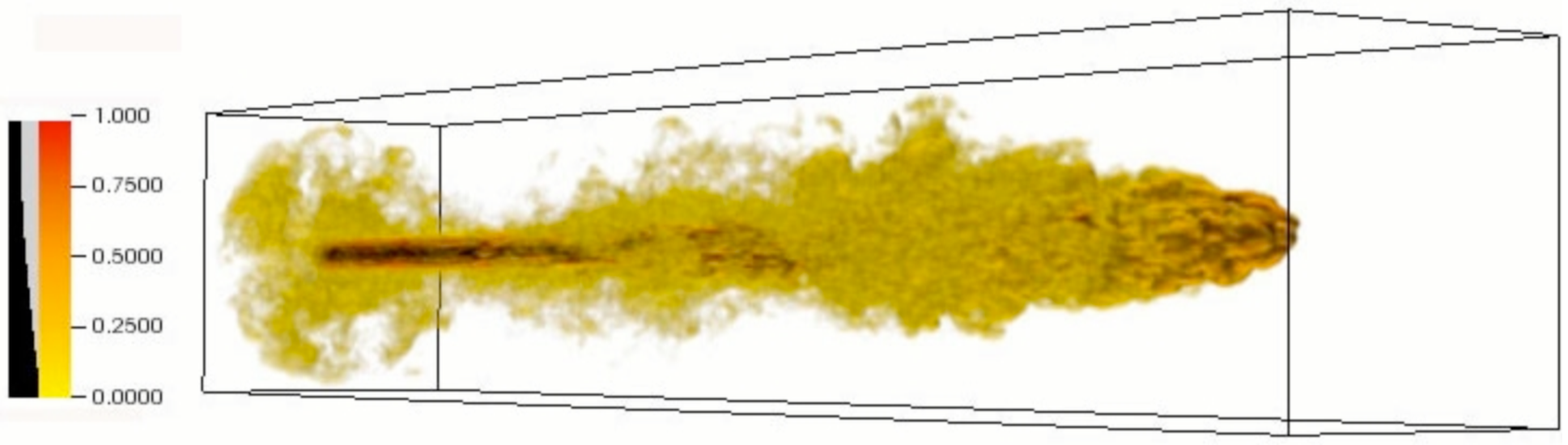} \\
\includegraphics[scale=0.5]{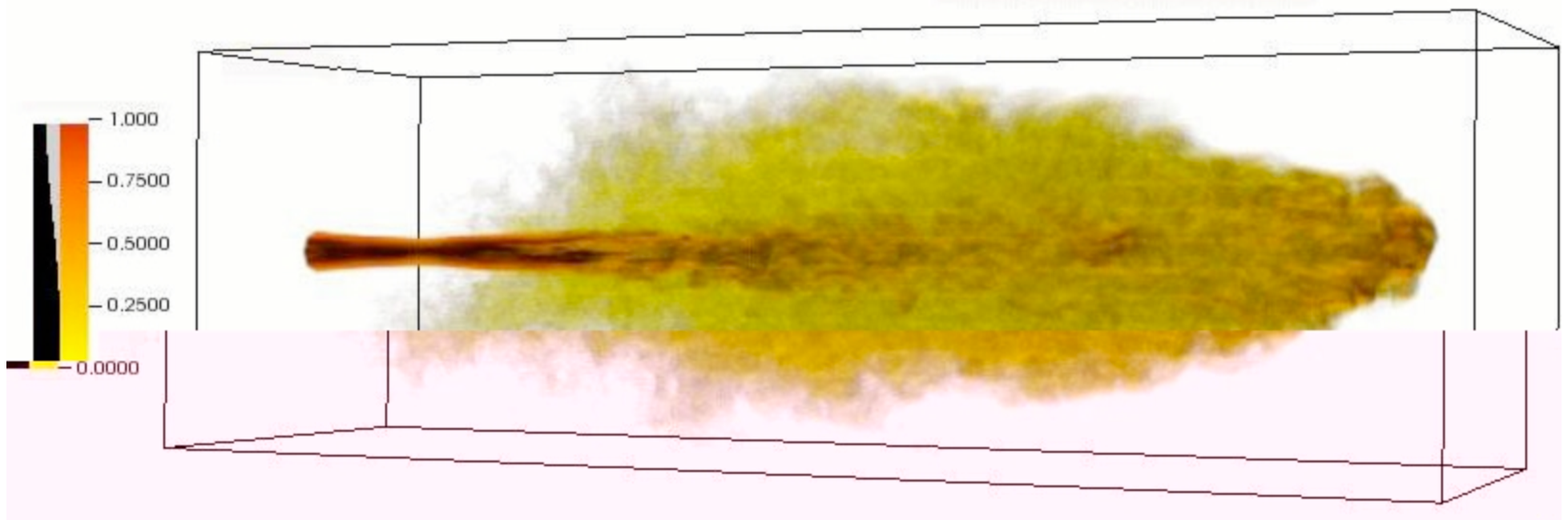} \\
\includegraphics[scale=0.5]{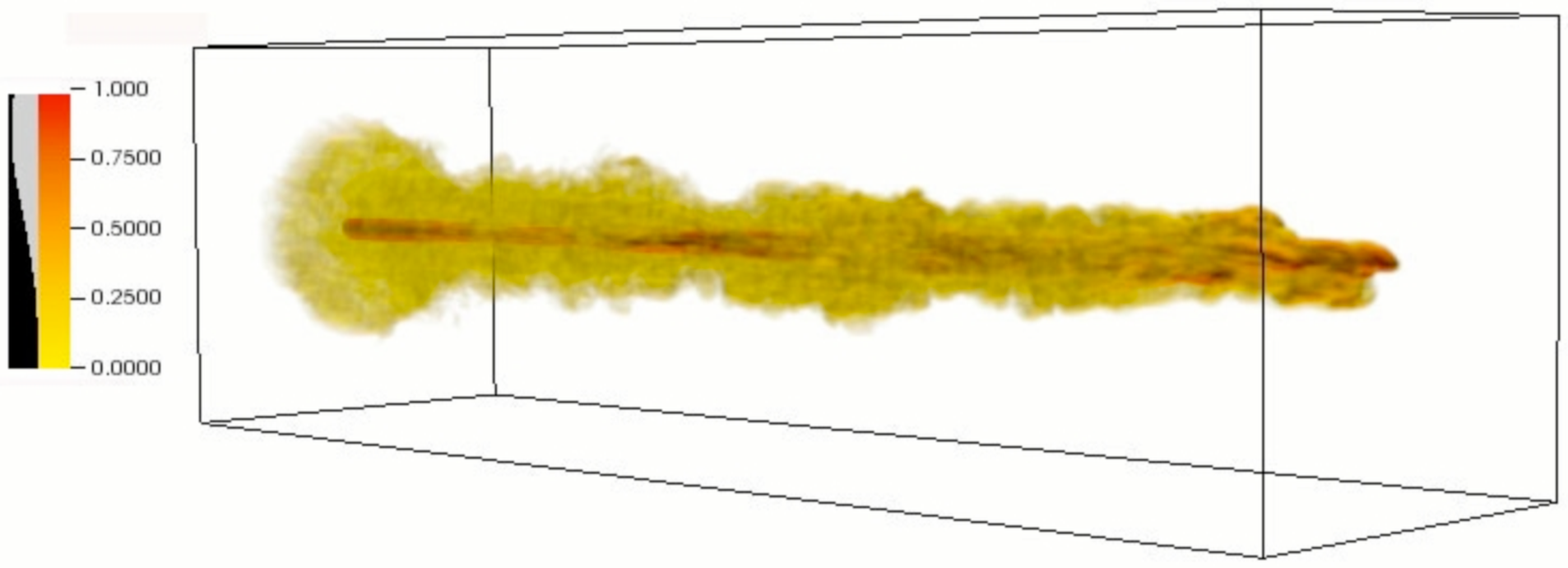} 
\end{tabular}
\caption{\footnotesize Volume rendering of the tracer 
         distribution for case A at $t=240$ (top panel), B at 
         $t=760$ (central panel) and E at $t=150$ (bottom panel). Dark
       colors are due to the opacity of the material in  the rendering procedure}
\label{fig:tracer}
\end{figure*}
With the choice of these elements we can cover a complete set of combinations
of Mach numbers and density ratios (cases A, B, D and E). Moreover, case C represents a lower
resolution version of case B, and a comparison between the two can provide indications 
on the role played by numerical resolution. In agreement with observations, we consider  only jets with density much 
lower than the external medium; on the other hand high density jet propagate quasi ballistically, 
do not show instabilities and mixing

As we shall discuss below, the main parameter
governing the jet behavior and the entrainment properties results to be the density ratio $\eta$,
while the Mach number plays a minor role even though it contributes to the determination 
of the KHI modes linear growth rates. Our discussion will therefore refer to cases A and B, as 
representatives of low and high $\eta$ values. Our results show that case D has a behavior 
very similar to case B, for this reason we will not discuss it in detail. Case E will be instead
properly discussed in order to show what can be the effects of a higher Mach number.   

Fig. \ref{fig:tracer} shows the volume rendering of the tracer distributions 
for case A (top panel), B (mid panel) and E (bottom panel) when the 
evolutions have reached fully nonlinear stages. Jet material is dark brown, 
external medium material white, level of mixing corresponds to the scale of orange.
From these figures one can gain a first qualitative view about the
role of the parameters on the evolution and structure of the jet: 
in cases A and E the jets propagate straight with moderate and 
low dispersion of jet particles, whereas in case B the jet has,
at least partially, lost its collimation and the particle dispersion
is considerably larger. 

\begin{figure*}
\resizebox{\hsize}{!}{
\begin{tabular}{c}
\includegraphics[]{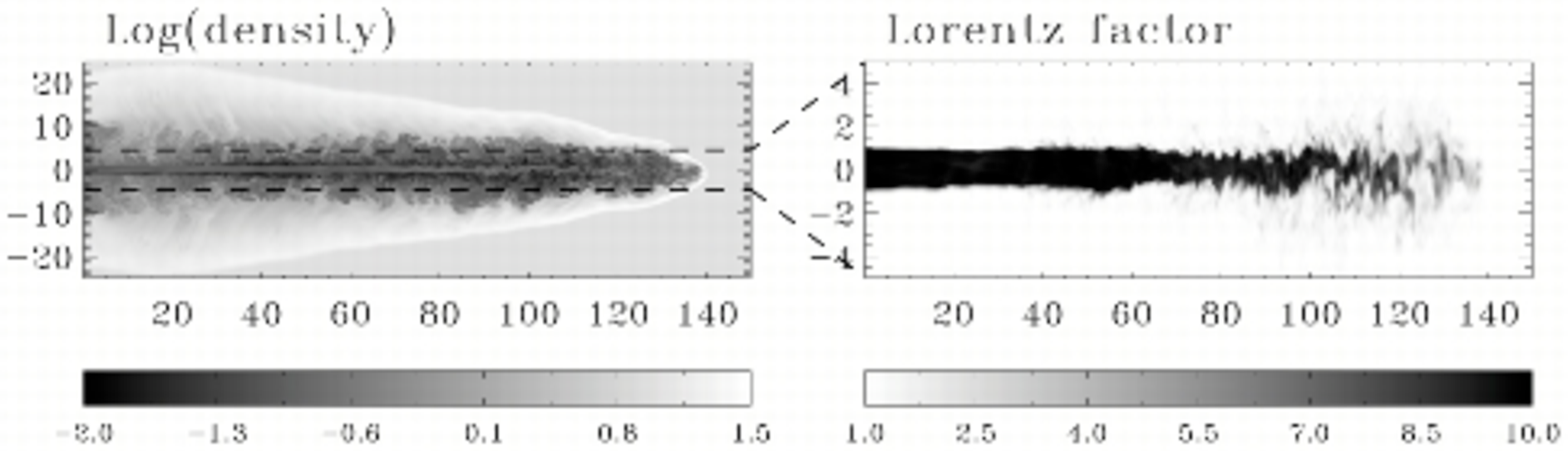} \\
\includegraphics[]{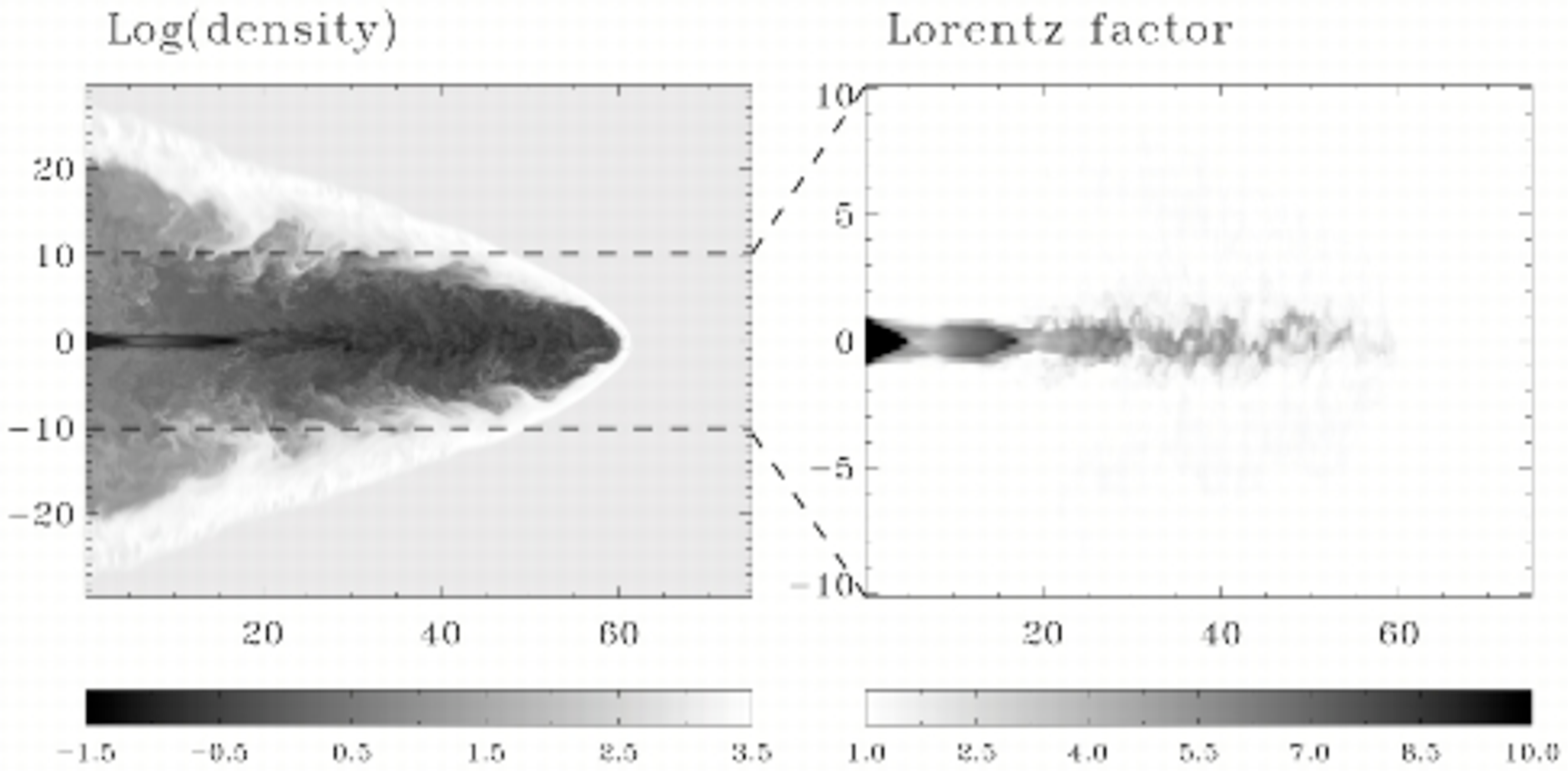} \\
\includegraphics[]{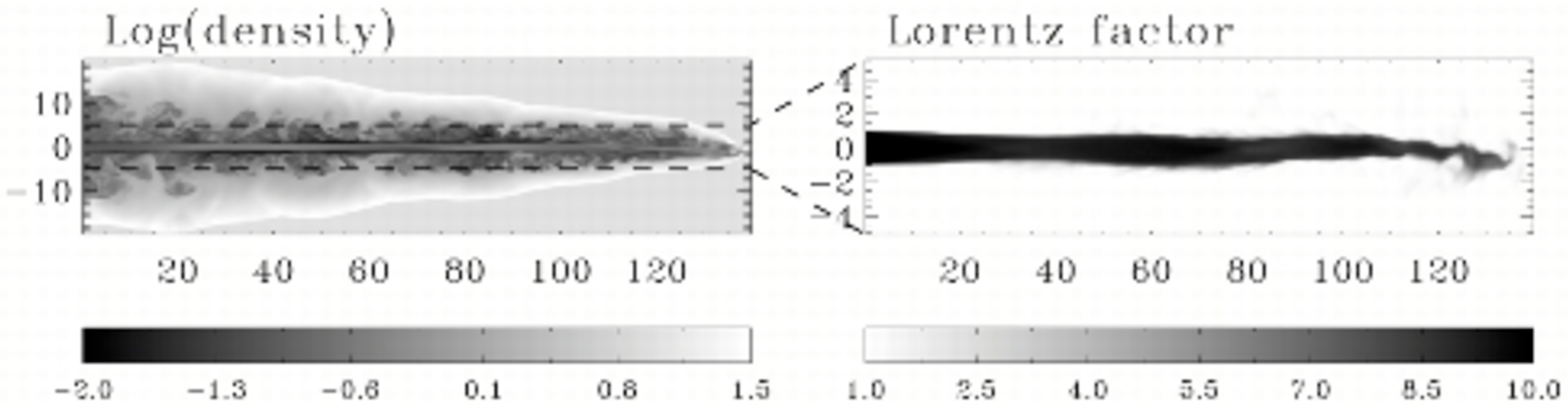} 
\end{tabular}
}
\caption{\footnotesize Two-dimensional longitudinal cut in the 
         $xy$ plane of the density distribution for case A at 
         $t=240$ (top-left panel), case B at $t=760$ (central-left panel) 
         and case E at $t=150$ (bottom-left panel), and Lorentz 
         $\gamma$ distribution in the central part of the
         domain, not in scale (the corresponding  right panels).}
\label{fig:dens_gamma}
\end{figure*}

This is more clearly represented by the 2-D cuts in the $x, y$ plane (at  $z=0$) 
of the density and Lorentz factor distributions shown in Fig. \ref{fig:dens_gamma}.
From the density (left) panels one can note that in both cases A and E
the jet seems to be weakly affected by the perturbation growth and entrainment. This is not 
the situation for case B, where the beam structure is heavily modified 
by the growth of disturbances beyond $\sim 20$ jet radii. 
The distributions of the Lorentz factor (right panels) indicate again that
the perturbation slightly affects the system in case A after $\sim 100$ radii,
leaves the jet almost unchanged in case E, and strongly influences the 
propagation after $\sim 20$ radii for case B, where the maximum value 
of $\gamma$ has reduced approximately to half of its initial value.  
However, even in case B a well collimated high velocity component along the axis still displays.

\begin{figure*}
\includegraphics[width=\hsize]{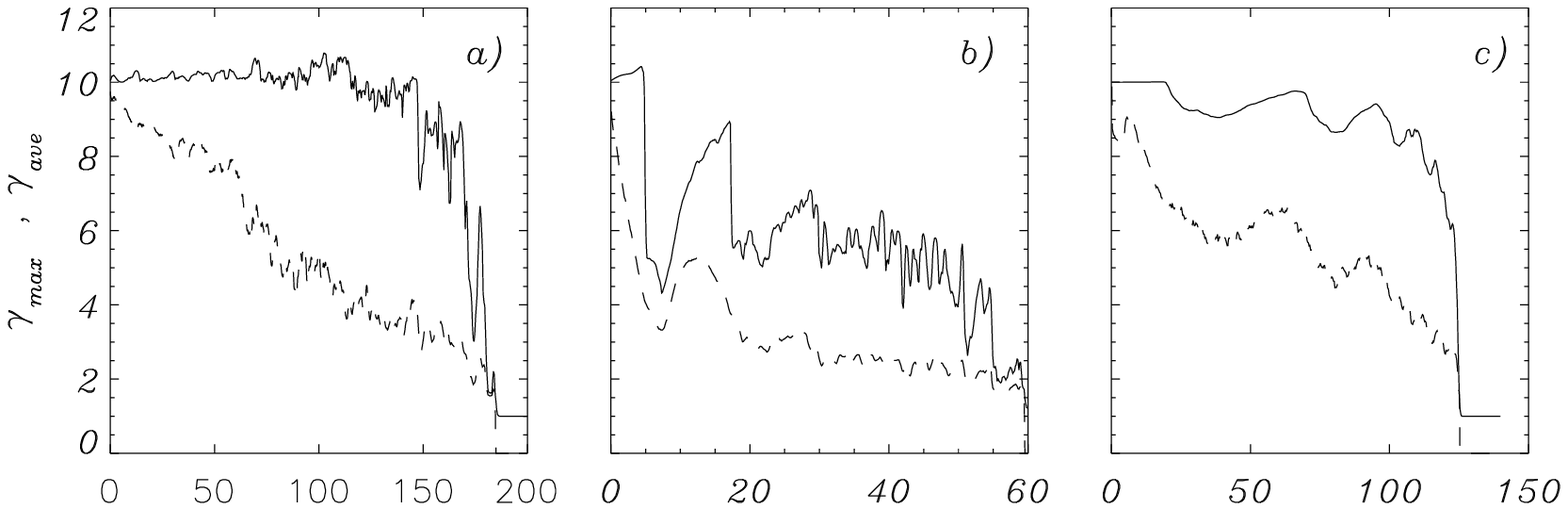}
\vspace{0.5cm}
\caption{\footnotesize Plots of the maximum value (solid line) and of an average
        value (dashed line) of $\gamma$ as functions of the longitudinal
        coordinate $y$ along the jet. The left panel refer to case A,
        the center panel to case B and the right panel to case E.}
\label{fig:gmax}
\end{figure*}

A quantitative estimate of the jet deceleration
can be obtained by plotting the Lorentz factor 
as a function of the longitudinal coordinate $y$.
Figure \ref{fig:gmax} shows the maximum value of $\gamma$ 
at constant $y$-planes together with its volume average, defined as
\begin{equation}
\gamma_{\rm av} = \frac {\int\gamma g(\gamma) \,dx dz}{\int g(\gamma) \,dx dz} \,,
\end{equation}
where  $g(\gamma)$ is a filter function to select the relativistic flows:
\begin{equation}
g(\gamma)=\Bigg\{
\begin{array}{ll}
1 & \qquad\textrm{for}\quad \gamma \geq 2  \,,
\\
0 & \qquad\textrm{for}\quad  \gamma < 2  \,.
\end{array} \Bigg.
\end{equation}
For case B one can see that the deceleration occurs both in 
$\gamma_{\rm max}$ (the flow velocity, although still relativistic, shows a strong decrease
from its initial value) and in $\gamma_{\rm av}$, which
indicates a global 
effect, whereas in cases A and E the central part of
the jet continues to be almost unperturbed and only thin external 
layers are decelerated.

\begin{figure}[!ht] 
\resizebox{\hsize}{!}{
\begin{tabular}{ccc}
\includegraphics[]{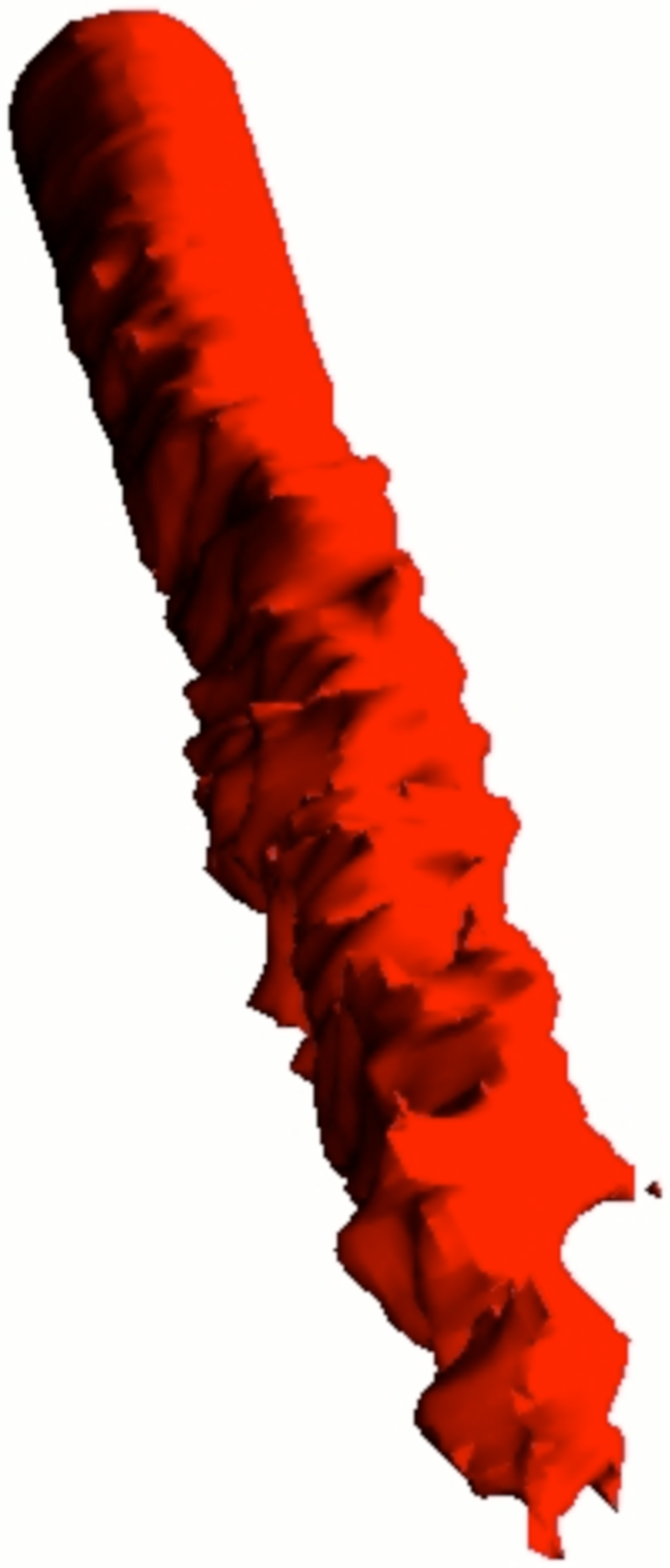} &
\includegraphics[]{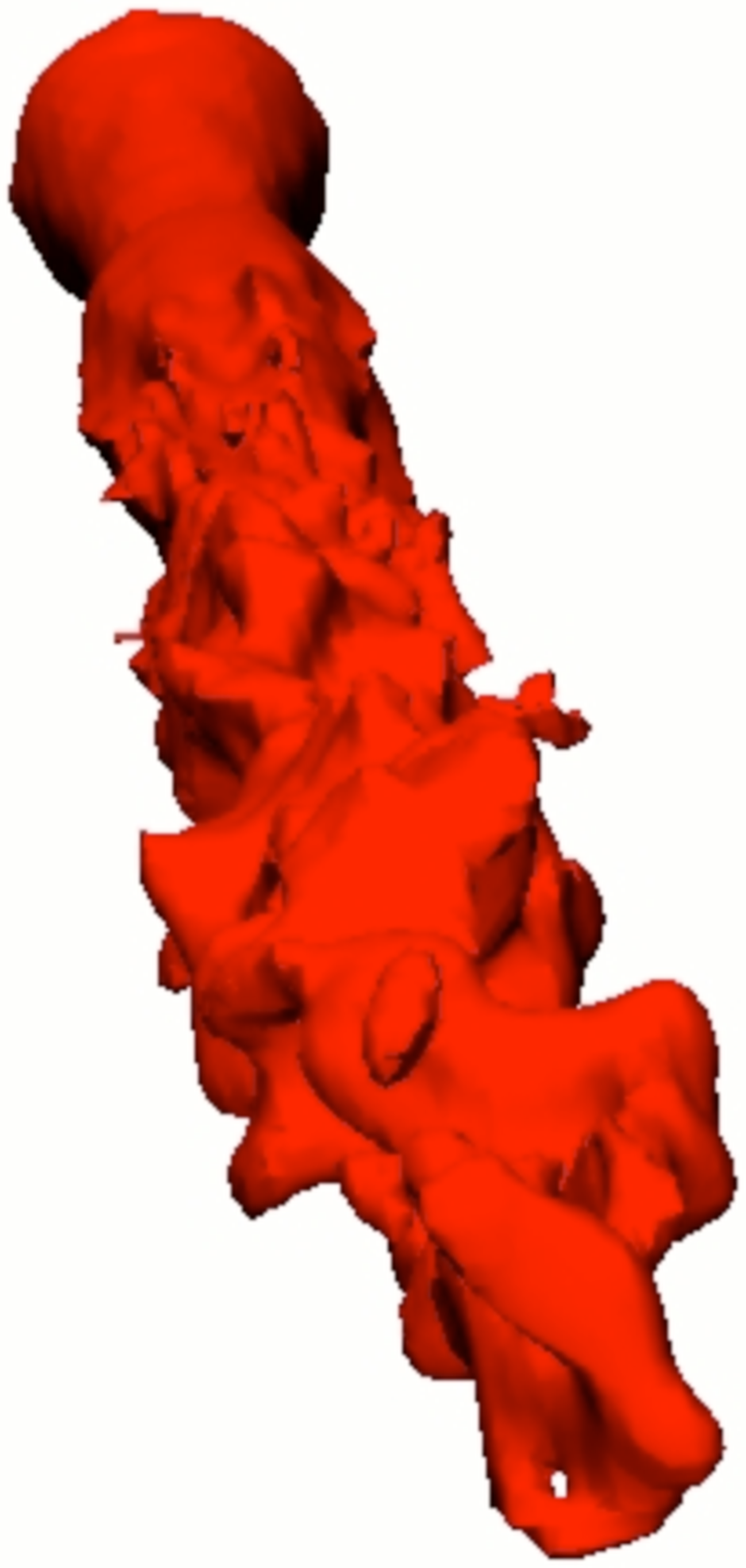} &
\includegraphics[]{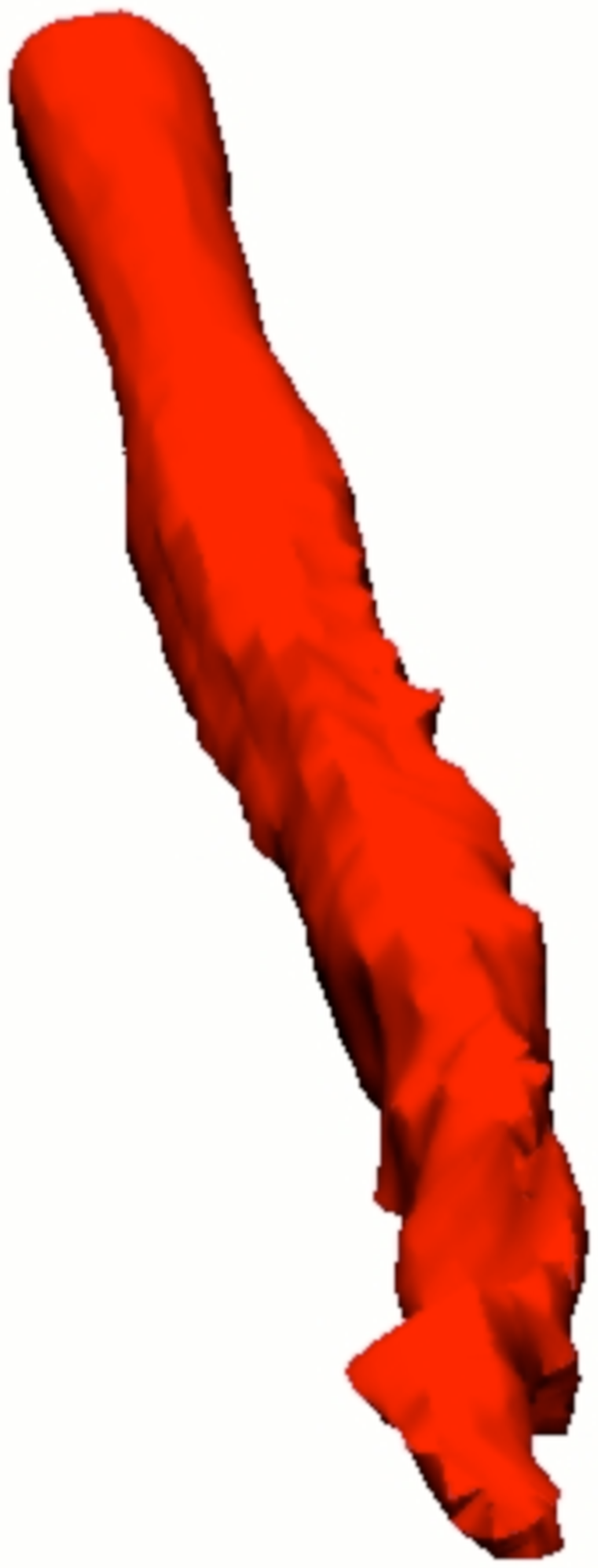} 
\end{tabular}
}
\caption{\footnotesize Image of the Lorentz factor distribution of the
         contour for case A (left panel, $\gamma=5$), case B 
        (central panel, $\gamma=3$) and case E (right panel, $\gamma=5$).}
\label{fig:lor}
\end{figure}

\begin{figure*}[!ht] 
\resizebox{\hsize}{!}{
\begin{tabular}{l}
\includegraphics[width=\hsize]{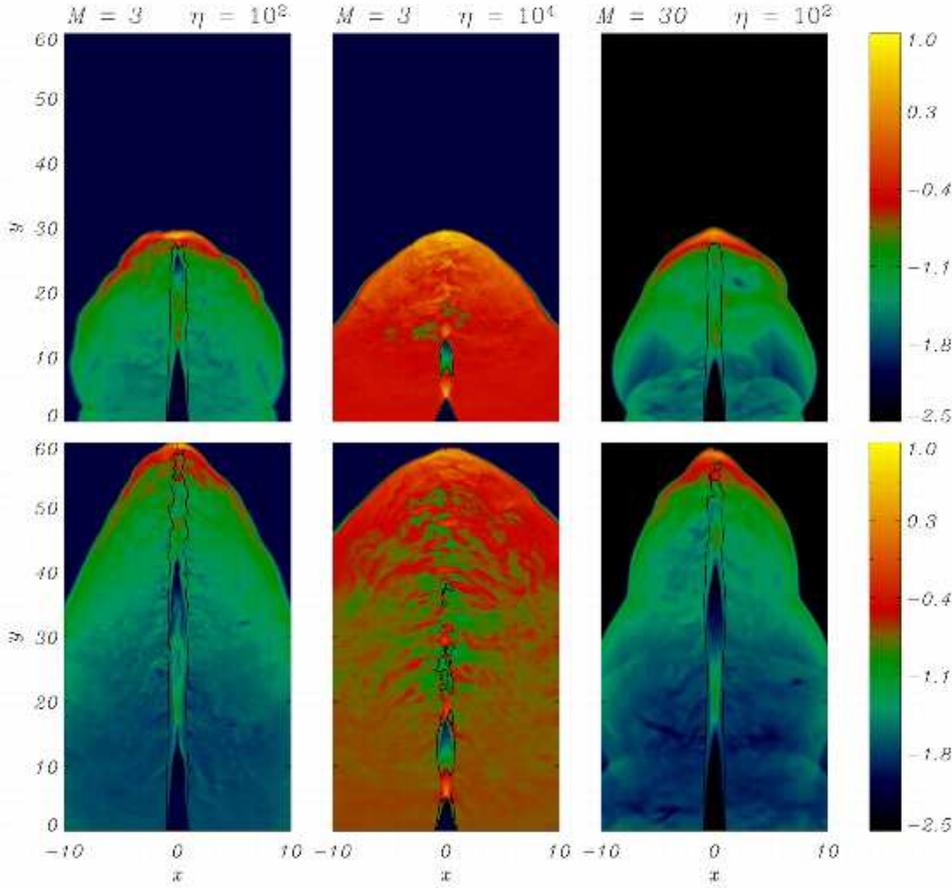} 
\end{tabular}
}
\caption{\footnotesize Transverse cuts in the $x-y$ plane of the pressure
         distribution with 
         superimposed the Lorentz factor contours. The left panels refer to case A, the
         middle panels to case B and the right panels to case E. The upper panels show the jet
         when it has reached a length of 30$r_{j}$ while, in the lower panels, the jet length is 
         60 $r_{j}$. 
         }
\label{fig:pressure}
\end{figure*}

\begin{figure}[!ht] 
\resizebox{\hsize}{!}{
\begin{tabular}{l}
\includegraphics[scale=0.57]{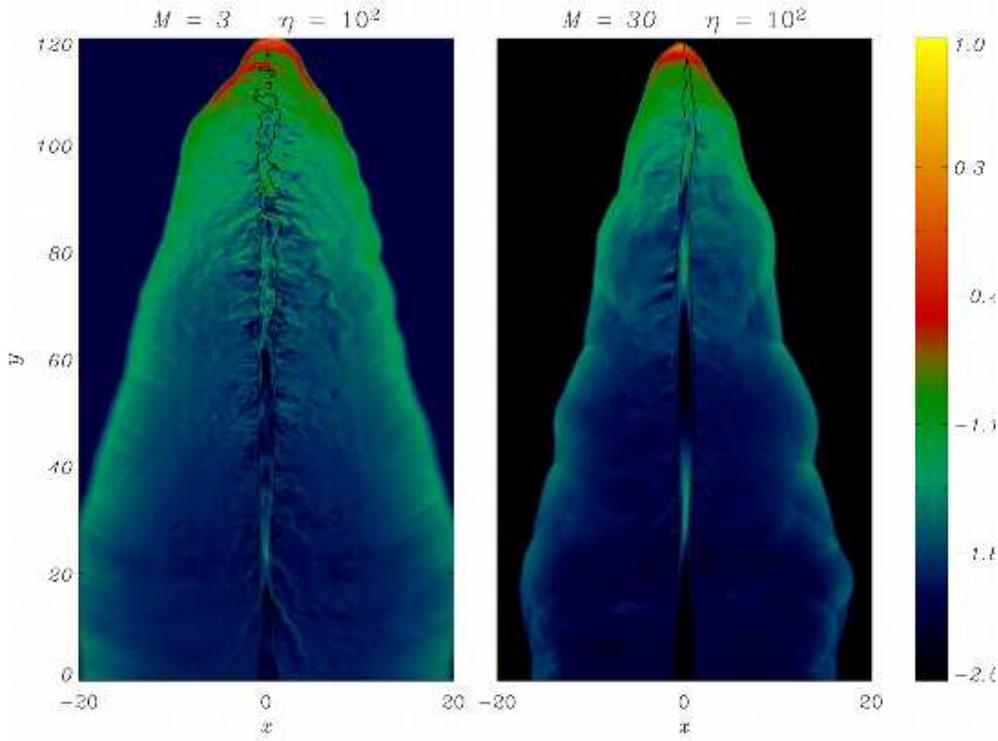}
\end{tabular}
}
\caption{\footnotesize Transverse cuts in the $x-y$ plane of the pressure
         distribution of the external material with 
         superimposed the Lorentz factor contours. The left panel refers to case A, the
         and the right panel to case E. The jet length is $120 r_{j}$ }
\label{fig:pressure1}
\end{figure}

\begin{figure*}[!ht] 
\resizebox{\hsize}{!}{
\begin{tabular}{l}
\includegraphics[width=\hsize]{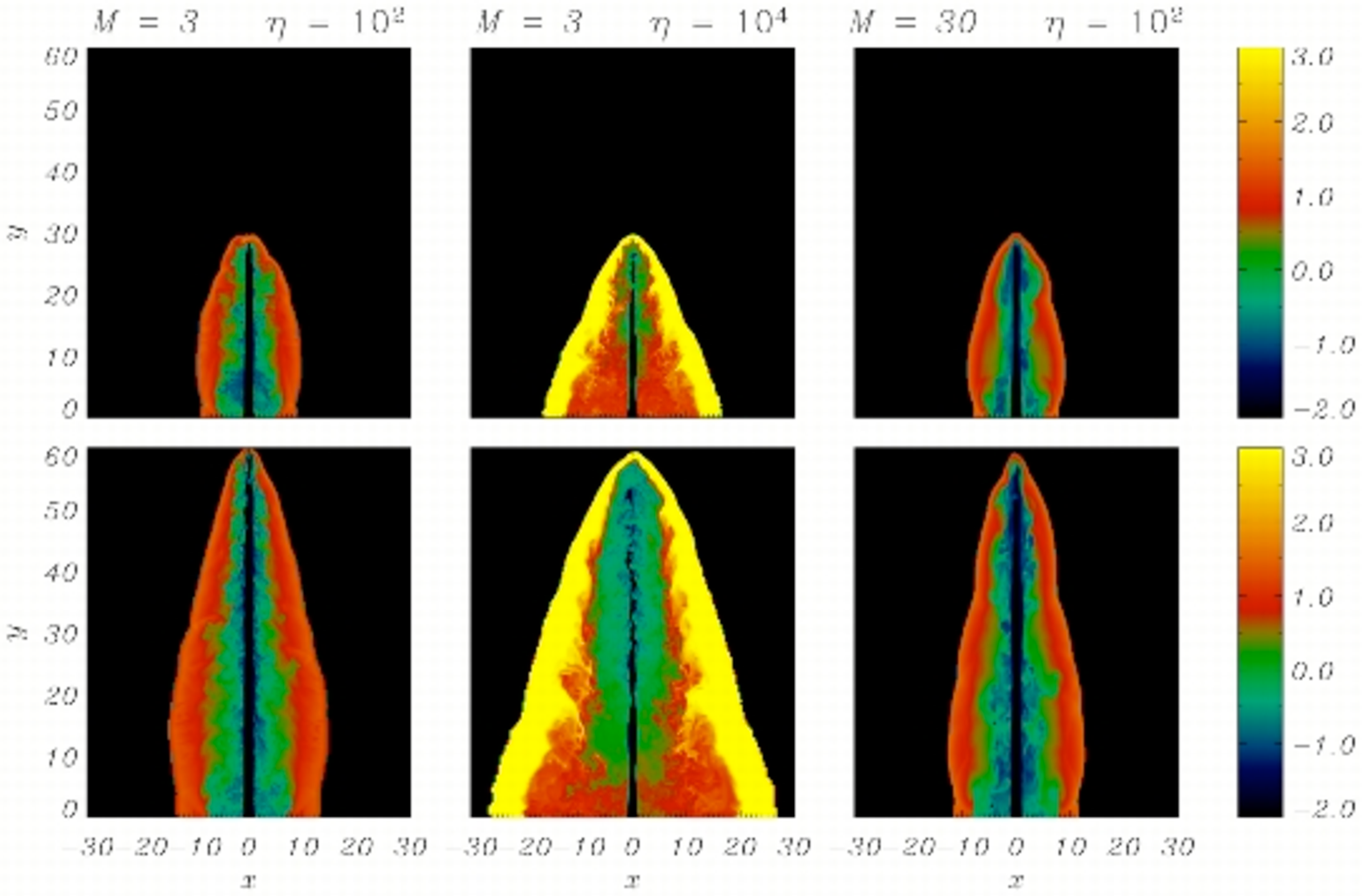} 
\end{tabular}
}
\caption{\footnotesize Transverse cuts in the $x-y$ plane of the density
         distribution of the external material.  The left panels refer to case A, the
         middle panels to case B and the right panels to case E. The upper panels show the jet
         when it has reached a length of 30$r_{j}$ while, in the lower panels, the jet length is 
         60 $r_{j}$.  }
\label{fig:ex_dens}
\end{figure*}

\begin{figure}[!ht] 
\resizebox{\hsize}{!}{
\begin{tabular}{l}
\includegraphics[scale=0.57]{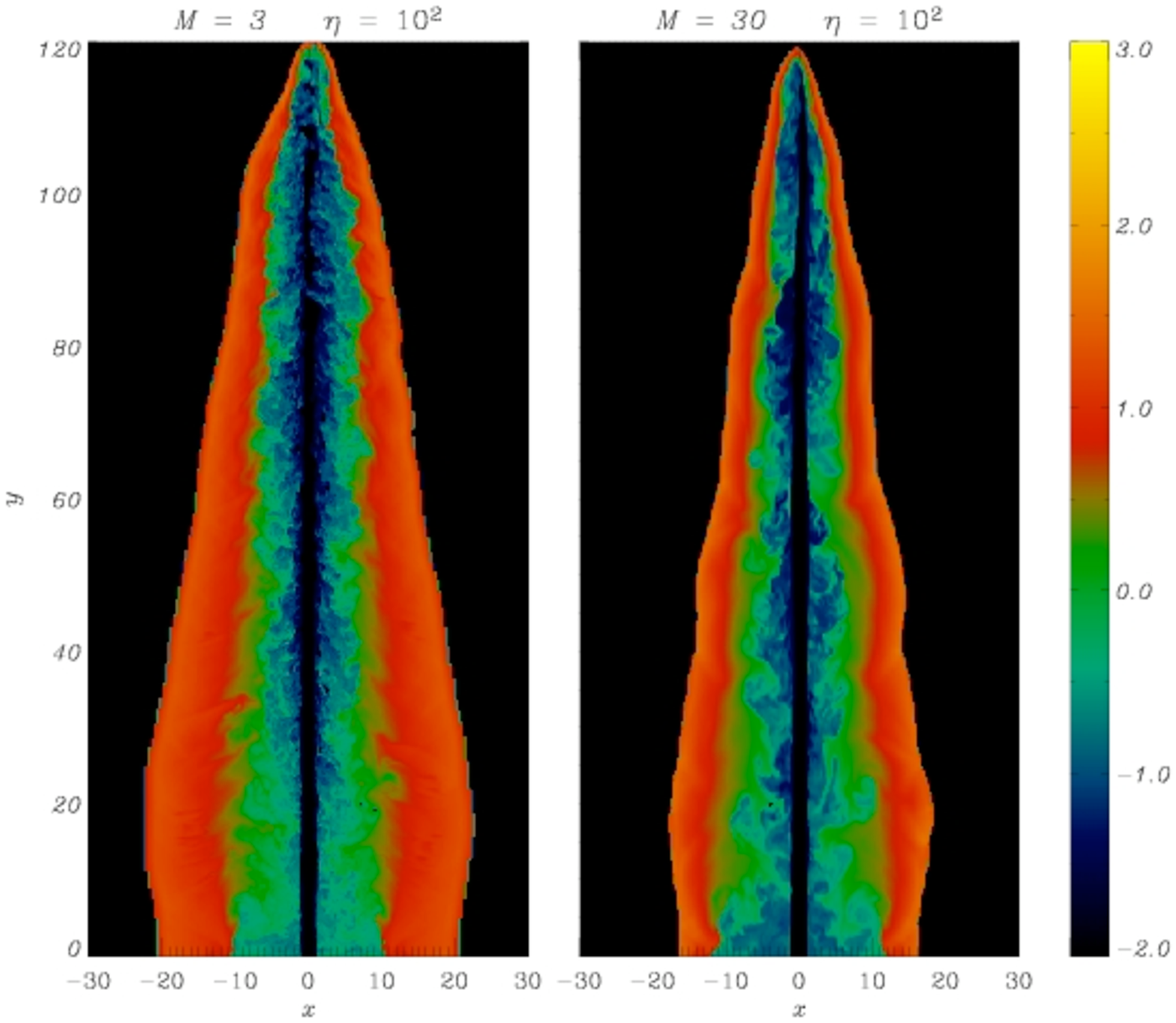}
\end{tabular}
}
\caption{\footnotesize Transverse cuts in the $x-y$ plane of the density
         distribution of the external material.  The left panel refers to case A, the
         and the right panel to case E. The jet length is $120 r_{j}$   }
\label{fig:ex_dens1}
\end{figure}

\begin{figure*}[!ht]
\resizebox{\hsize}{!}{
\begin{tabular}{l}
\includegraphics[width=\hsize]{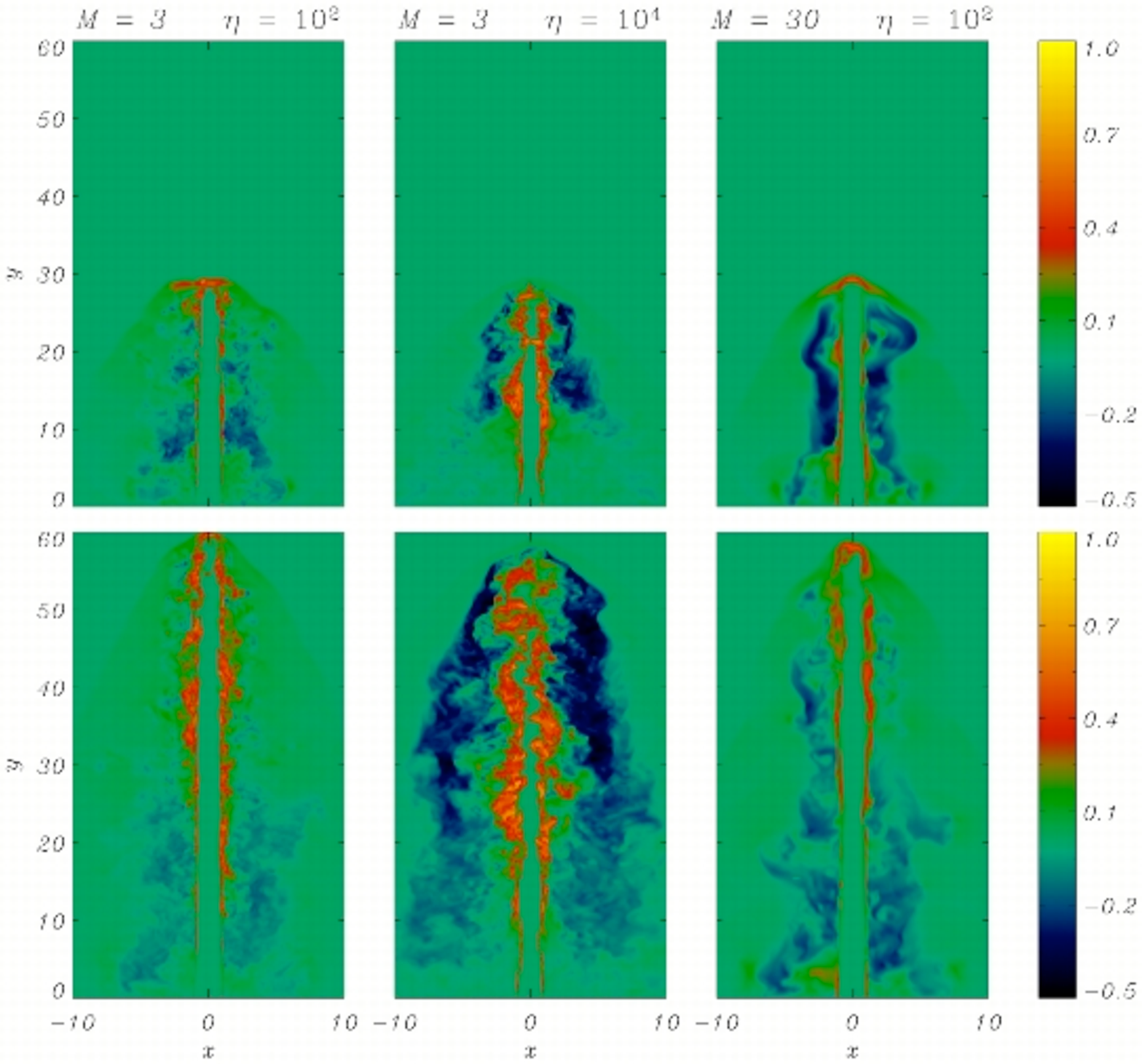} 
 \end{tabular}
}
\caption{\footnotesize Transverse cuts in the $x-y$ plane of the $v_{y}$ distribution 
         of the external medium.  The left panels refer to case A, the
         middle panels to case B and the right panels to case E. The upper panels show the jet
         when it has reached a length of 30$r_{j}$ while, in the lower panels, the jet length is 
         60 $r_{j}$.}
\label{fig:ex_beta}
\end{figure*}

\begin{figure}[!ht]
\resizebox{\hsize}{!}{
\begin{tabular}{l}
\includegraphics[scale=0.57]{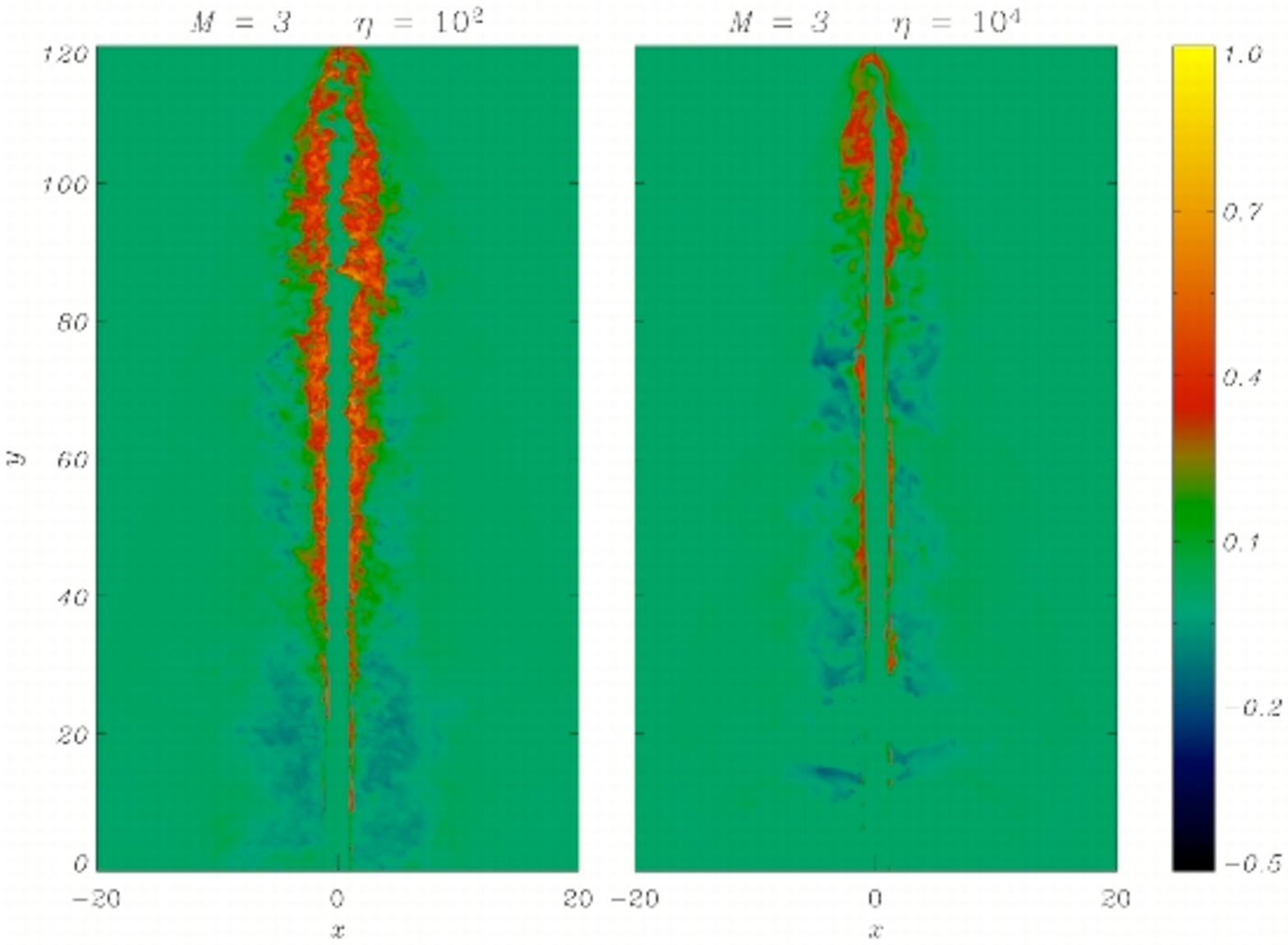}
 \end{tabular}
}
\caption{\footnotesize  Transverse cuts in the $x-y$ plane of the $v_{y}$ distribution 
         of the external medium.   The left panel refers to case A, the
         and the right panel to case E. The jet length is $120 r_{j}$  }
\label{fig:ex_beta1}
\end{figure}

As previously mentioned, the interaction between jet and surrounding
medium involves different phenomena, reflecting on the characteristics  
of the deceleration process: the growth rate of the perturbations, 
the type of perturbations that dominates the jet
structure, the possibility of mixing through the backflowing material
and the density of the ambient medium relative to the jet
material. Long wavelength modes will tend to produce global jet
deformations like jet wiggling more than mixing, while modes at
shorter scale may be more efficient in promoting the mixing
process. On the other hand, it is clear that the mixing with a denser
environment is more effective for the jet deceleration. One of the
differences between the two cases that we are considering is, in fact,
the type of modes that dominate the jet evolution: while case E exhibits
longer wavelength and lower $m$ modes with a highly prominent
helical distortion, in case B we observe shorter wavelength and higher $m$
modes. This behavior agrees with the linear analysis of
\cite{Hardee87} that predicts that wavelengths of the resonant
mode increases with the Mach number as $\lambda \propto \gamma M$.
An impression of the characteristic scales of the dominant modes can be
obtained from Fig. \ref{fig:lor}, where 3D contour images
of the Lorentz factor are shown.
The central panel, case B, shows that small scale structures
dominate the structure all along the jet, while case E (right panel) shows the 
dominance of the $m = 2$ helical mode. The left panel (case A), denotes a 
less clear cut situation possibly due to the effect of the superposition of 
different modes, with the presence of structures of scale smaller than
in case E.
 
So far we have compared our three simulations at
a given time. In Figs. \ref{fig:pressure}, \ref{fig:pressure1}, \ref{fig:ex_dens},
\ref{fig:ex_dens1}, \ref{fig:ex_beta} and \ref{fig:ex_beta1} we consider as well the temporal evolution of pressure, 
density and velocity,
comparing the jet structures at three different times when they reach 
approximately equal lengths. 
In Fig. \ref{fig:pressure} we show cuts in the $(x,y)$ plane of the
pressure distribution with superimposed contours of the Lorentz
factor. The figure displays the formation of the typical series of
regular biconical structures \citep{KomissarovFalle98} produced by the
overpressured cocoons squeezing the jet. Case B shows the highest
pressure and has shocks of smaller obliquity. From the contour levels of the
Lorentz factor we see once more that cases A and E keeps the initial 
value of 10 up to the jet head. Fig. \ref{fig:pressure1} shows the jets
when they have reached a length of 120 radii; here we display only cases A and E since,
for case B, the jet head velocity is very low and the simulation was ended at a jet
length of 60 radii. In the figure, we start seeing differences
between case A, more perturbed, and case E, that shows almost no sign of
interaction between the jet beam and the cocoon. Differences
between cases A and E  were already evident in Fig. \ref{fig:lor},
that showed the presence of small scale perturbations more favourable
to mixing in case A and of only a long wavelength helical distortion
in case E. 

On the contrary, for case B that shows the
strongest deceleration, a first decrease of the Lorentz
factor occurs at the first biconical shock (see also
Figs. \ref{fig:dens_gamma} and \ref{fig:gmax}), the reacceleration following
the shock however does not lead to a complete recovery of the initial
value because the jet beam starts to interact with the cocoon and to
lose momentum. We can note that, in the relativistic case,  an
increase of  the jet pressure to resist squeezing by the overpressured cocoon would result in
an increase of the jet energy flux (due to the relativistic
contribution of pressure to inertia) and in a further increase of the
cocoon pressure. Thus, in the relativistic regime, the formation of
biconical shocks appears  to be even more unavoidable than in the nonrelatistic regime.

The properties of the entrainment process can be better understood
by looking at Figs \ref{fig:ex_dens}-\ref{fig:ex_beta1}.  
In Fig. \ref{fig:ex_dens} and \ref{fig:ex_dens1} we show cuts in the $(x,y)$ plane of the 
density distribution of the external medium (derived from the tracer
distribution)  for the three cases considered, 
at different times. This figure shows the ability of the external
medium to penetrate the contact discontinuity and mix with the jet beam.
We see that in B (central panel) the jet is surrounded by denser 
(green-red) external material in most of its length, while in cases A (left
panels) and E
(right panels) the medium around the jet is always lighter  (blue-green). 
Comparing the different evolutionary stages (see Figs \ref{fig:ex_dens} and \ref{fig:ex_dens1}), 
we also observe a slight decrease in density. 
The effect of the jet on the speed of entrained matter is shown in 
Figs. \ref{fig:ex_beta} and \ref{fig:ex_beta1}, where we display cuts of the distribution of 
the external medium velocity component along the jet axis. We note that the
volume of the accelerated medium (red) is considerably higher in case B
than in cases A and E, consistently with a substantially stronger 
instability. Entrainment in case B starts to be
effective already at a length of 30 radii, when it is almost absent in
the other cases. At a length of 60 radii, apart from the dominant case
B, we start observing differences between case A and case E with a
stronger entrainment effect of the first. This difference becomes even
more prominent at a length of 120 jet radii (see Fig. \ref{fig:ex_beta1}). From this figures we can
also observe the formation of the backflow (blue region), more
prominent in case B and almost absent in cases A and E. 
\section{Entrainment properties}
\label{sec:entr}
%
%
%
%
%
%

We now discuss in detail the entrainment properties of the jet of case B ($M = 3$,
$\eta = 10^4$). We begin examining the distribution in Lorentz
factor and velocity of the jet and external material at $t = 600$
(at the end of the simulation).



\begin{figure*}
\includegraphics[width=\hsize]{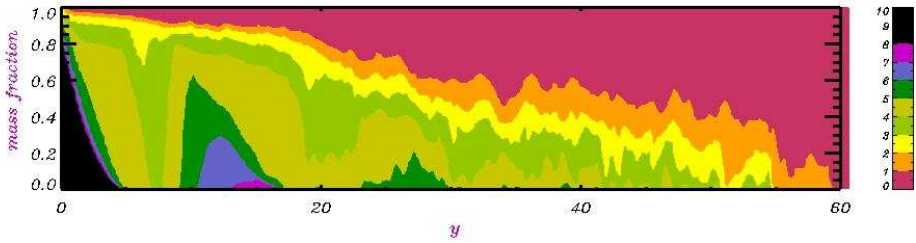}
\caption{\footnotesize Distribution of the jet mass fraction moving at a given
         $\gamma \beta$, when the
         jet length is $60 r_{j}$}
\label{fig:mjetd}
\end{figure*}

In Fig. \ref{fig:mjetd} we show the distribution of the jet mass fraction
that is moving at a certain value of $\gamma \beta$ at a given time. We use the four-velocity
(instead of the Lorentz factor) to avoid compression of the
scale close to $\gamma = 1$, i.e. at low velocities. The color scale
on the right gives the corresponding value of $\gamma \beta$ for each
color. As an illustrative example, one can see that at $y=20$, $\sim 20\%$ of the
jet mass (per unit length) moves at $4 < \gamma \beta <
5$, $\sim  40 \%$ moves at $3 < \gamma \beta < 4$, $\sim 17 \%$ moves 
at $2 < \gamma \beta < 3$, $\sim 5 \%$ moves at $1 < \gamma \beta < 2$
and finally $\sim 18 \%$ moves at $0.1 < \gamma \beta < 1$. In this
image one can clearly recognize the effect of reacceleration after internal shocks driven in the
beam by the high pressure cocoon, particularly close to the inlet region. The
central region of the jet never decelerate below $\gamma = 4$, while
in the external layers of the jet we observe the formation of an
expanding region of low velocity material ($\gamma \beta < 1$).    

\begin{figure*}
\begin{tabular}{ccc}
\includegraphics[scale=0.33]{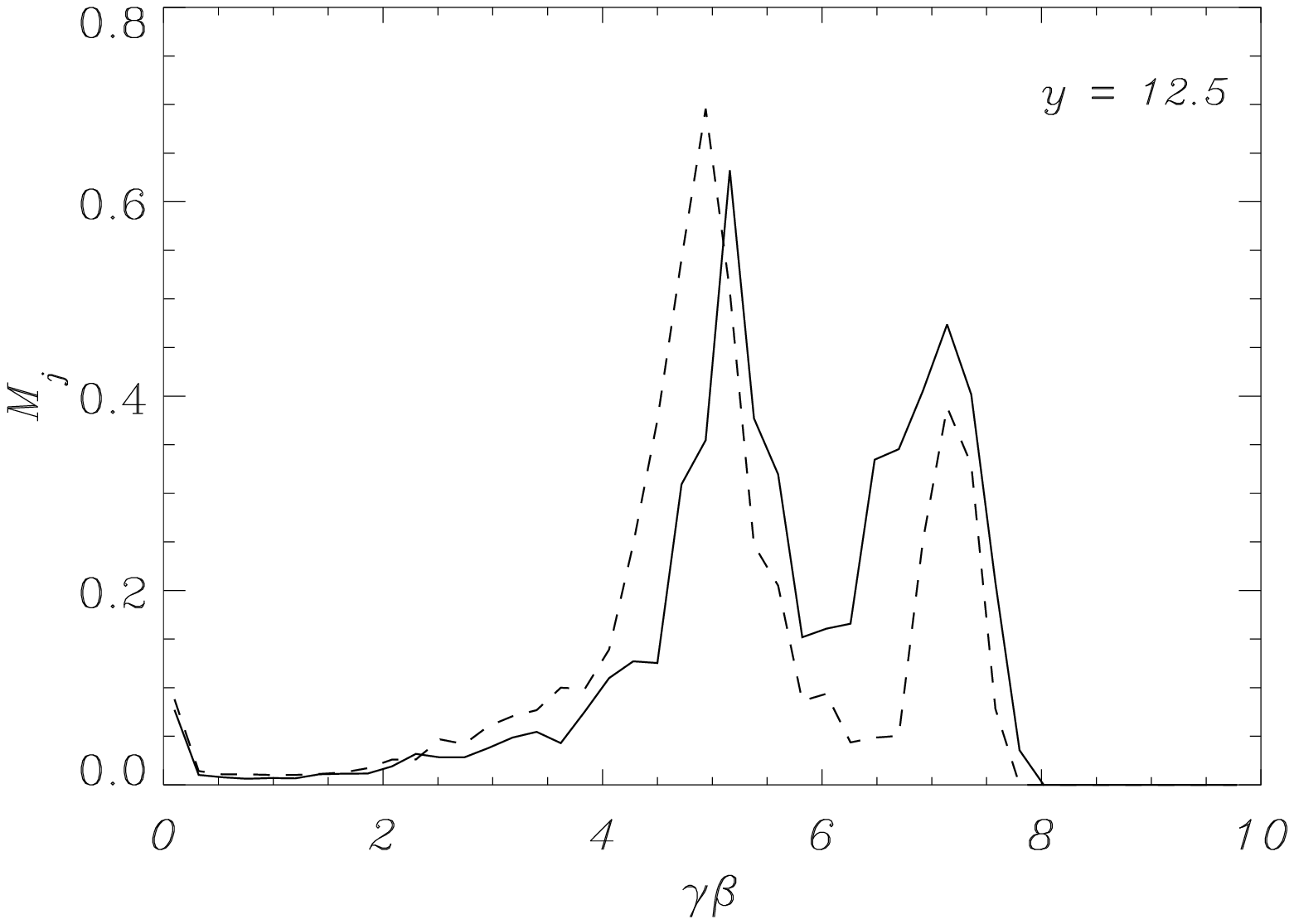} &
\includegraphics[scale=0.33]{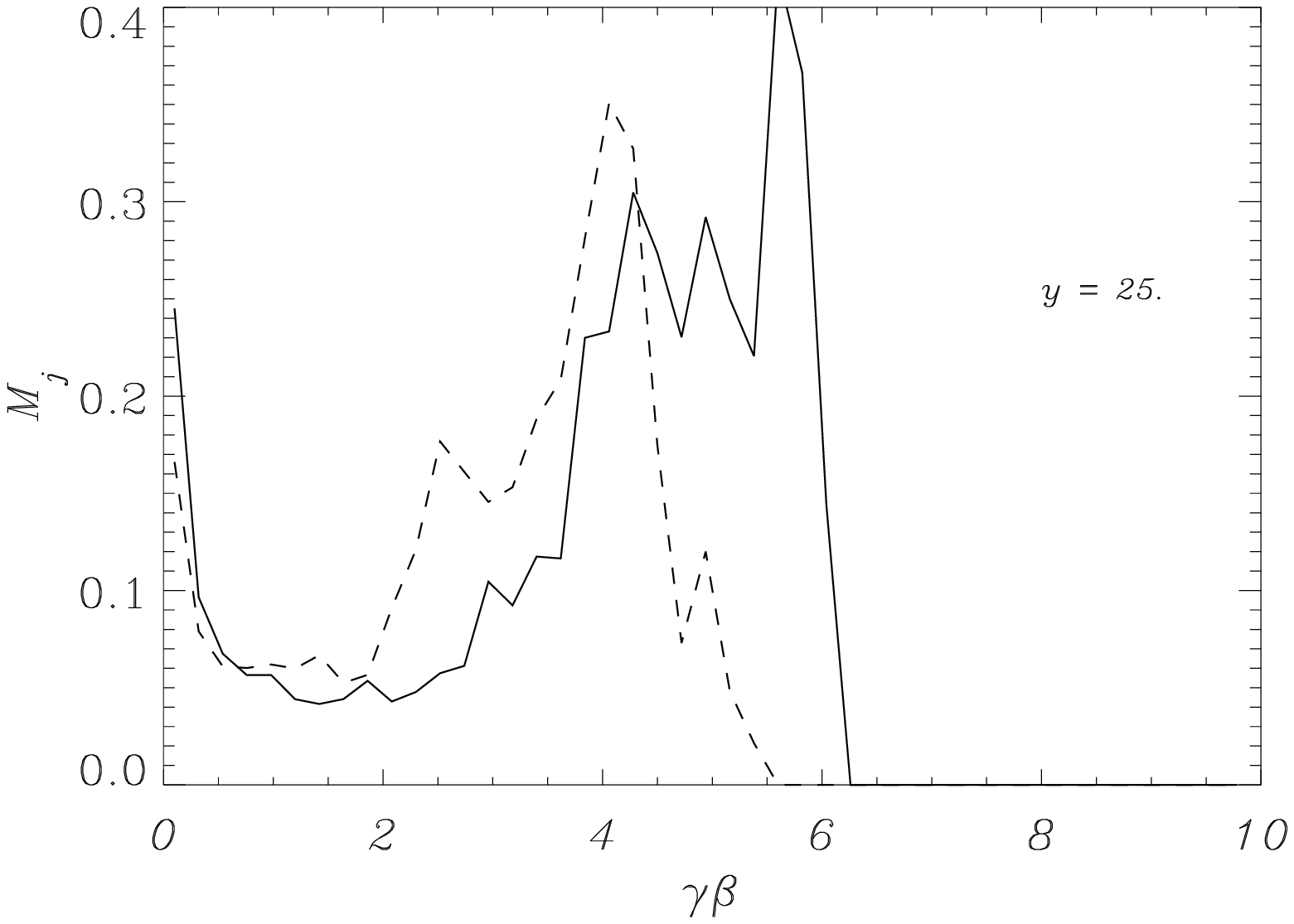} &
\includegraphics[scale=0.33]{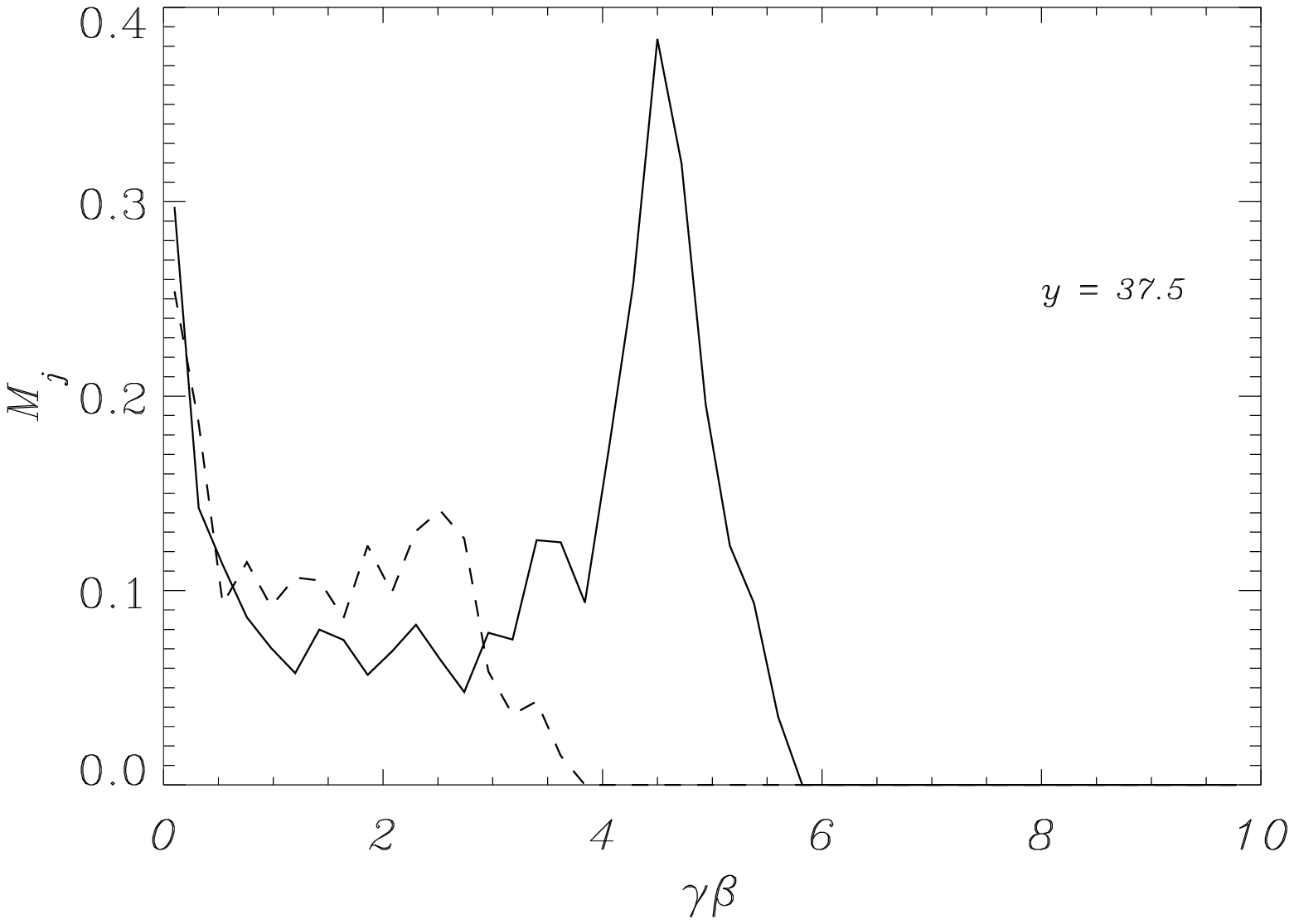}
\end{tabular}
\caption{\footnotesize Jet mass distribution as a function of $\gamma \beta$ at the 
         locations $y=12.5, 25, 37.5$ at $t=400$ (dashed line)
         and $t=600$ (solid line)}
\label{fig:mdis1}
\end{figure*}

Additional details are illustrated in Fig. \ref{fig:mdis1},
where the jet mass distribution is plotted as a function of $\gamma \beta$
at three different values of $y$ for two different instants.
The three panels, from left to right, refer respectively to
$y = 12.5$, $y = 25$ and $y = 37.5$, at $t = 400$ (dashed line)
and $t = 600$ (solid line).
At the first position ($y=12.5$) the differences between the two
times is relatively small, while at the two other locations we have a
significant increase of the material moving at high $\gamma$. The jet
in its first part has reached a quasi-steady configuration, while at 
larger distances from the source, after an initial deceleration phase, 
a well collimated high velocity core appears. 
Looking at the  velocity structure we see that, at the beginning
($y=12.5$, see
the left panel of Fig. \ref{fig:mdis1}), the jet is predominantly at
high values of $\gamma \beta$, even though the deceleration mechanism
starts to be effective, as shown by the small peak at $\gamma \beta
\sim 0.2 $.
The effects of entrainment become much more evident at larger
distances (see the two rightmost panels in Fig.\ref{fig:mdis1}), where
we observe the formation of two 
sharp peaks, one at high $\gamma \beta$ and the other at velocities 
$\gamma \beta \sim 0.2 $, with considerably less material at intermediate 
velocities. 
This implies that the jet structure has two well defined velocity 
components with a steep shear layer between them. 

\begin{figure*}
\includegraphics[width=\hsize]{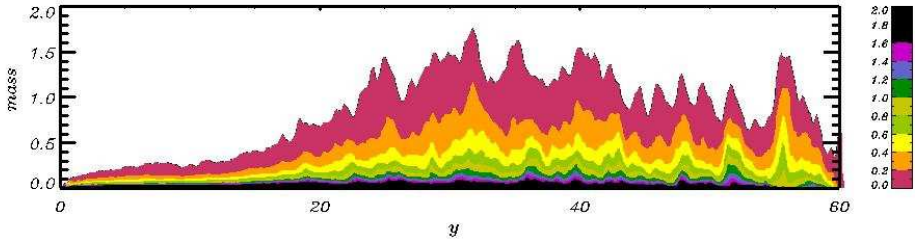}
\caption{\footnotesize The same as in Fig. \ref{fig:mjetd} for the external medium
         and normalized with $M_j(0)$}
\label{fig:mextd}
\end{figure*}

The distribution of external mass, analogous to Fig. \ref{fig:mjetd} is
also plotted , in Fig. \ref{fig:mextd}.
In this case all quantities are normalized to the jet mass injected
in the domain per unit time. 
%

The figure show that most of the external material is moving at quite
low velocities, i.e. $0.1 < \beta < 0.3$. A similar analysis 
on the backflow reveals that most material moves with velocity in 
the range $-0.3 < \beta < -0.1$, with a strongly dominant contribution 
given by the external medium.
Higher velocity fluid elements with velocities up to $-0.8$ can be found
in the front regions of the jet again with a dominant contribution 
by the ambient medium.

We conclude this section with a short evaluation of the effect of
numerical resolution. Case C has the same parameters as case B, but
only 12 points per jet radius, compared to 20 points per jet radius of
case B. Comparing the results obtained at these two different
resolutions, we have obtained that the general effect on the entrainment
is to increase its efficiency when we increase the
resolution. A detailed comparison shows that, for example, the average
Lorentz factor decreases by about 15\% in the higher resolution
run. This effect can be interpreted observing that the most effective
modes in term of entrainment are those at a shorter wavelength, that
are under-resolved in a low resolution run.   

\section{Astrophysical implications}
\label{sec:implic}
%
%
%
%
%
%
Recently several authors, e.g. \citep{Chiaberge00, 
Piner04, Giroletti04}, in order to explain observational properties 
of FR-I radio sources and their beamed counterparts (BL Lac objects), 
have proposed that they are produced by jets characterized by
a velocity structure in which an inner core
maintains a highly relativistic velocity and is surrounded by
material that has been slowed down by the interaction with the
ambient medium.  A structure of this type has been called ``spine-layer".
The appearance of a jet with a spine-layer configuration is different 
when viewed at different angles. 
In fact, the two velocity components have different
Doppler factors and the spine dominates the emission when the jet
is observed at small angles with respect to the line of sight 
(BL Lac objects with strong Doppler boosting), 
while the prevailing contribution at larger angles is due to the entrained 
layer at low Lorentz factors (FR-I radio sources)

In our calculations, a "spine-layer" velocity structure has been
obtained self-consistently as the result of a well defined
physical process, i.e. the interaction of the outer jet layers with
the ambient material, driven by jet instabilities. 
In particular we have found that, in the strongly underdense 
case $\eta=10^4$, the jet acquires a velocity structure in which the inner core
maintains a highly relativistic velocity and is surrounded by
material that has been slowed down by the interaction with the
ambient medium.
Therefore we attempt a comparison of radio maps 
constructed from the simulated jets with observations of FR-I jets. 
To this purpose, we compute synthetic maps by integrating the synchrotron 
emissivity along the line of sight.
For the sake of simplicity we assume the emissivity to be proportional 
to the proper density  of the jet material multiplied by the appropriate boosting or 
deboosting factor, i.e. by the quantity
\begin{equation}
\epsilon(x,y,z) =  \left[ \frac{1}{\gamma (1- \beta \cos \theta)} 
\right]^{2+\alpha} \;,
 \label{eq:emis}
\end{equation}
where $\alpha$ is the spectral index of the radio flux 
(we take $\alpha=0.5$).
From this calculation we exclude the material at rest 
accumulated at the base of the jet as a result of the 
reflecting boundary condition imposed at $y = 0$. 
The presence of this material is 
an artifact due to the imposed equatorial symmetry.   

\begin{figure} 
\includegraphics[width=\hsize]{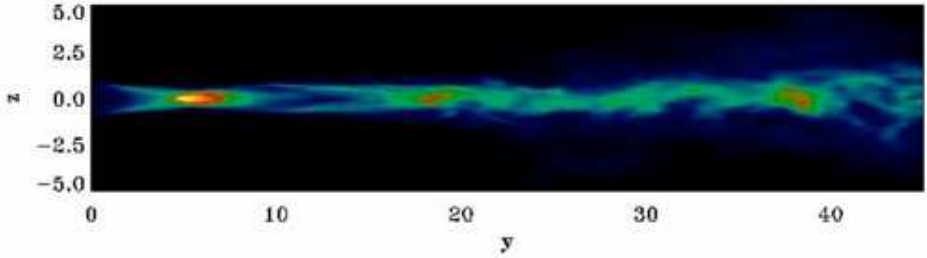} 
\caption{\footnotesize Synthetic map for the jet of case B with an inclination of
         $20^\circ$ with respect to the line of sight.}
\label{fig:map20}
\end{figure}

\begin{figure} 
\includegraphics[width=\hsize]{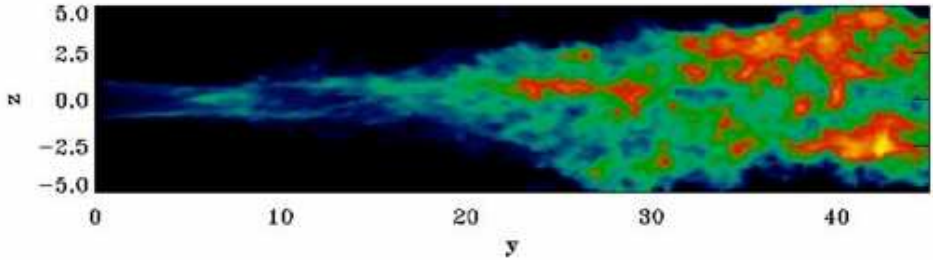} 
\caption{\footnotesize Synthetic map for the jet of case B with an inclination of
         $60^\circ$ with respect to the line of sight.}
\label{fig:map60}
\end{figure}

The synthetic maps for the strongly decelerated jets of case B are plotted 
in Figs. \ref{fig:map20} and \ref{fig:map60} for inclination angles of the jet axis to
the line of sight of $20^\circ$ and $60^\circ$ respectively. The maps show a
region with equal projected lengths of $45$ jet radii, excluding the jet head,
and therefore correspond to a longer jet in the small angle case. 
In the small angle case, the relativistic jet $\gamma \sim 4\div 5$ core emission 
dominates. Emission is present along the whole
jet and some knots can be recognized, the most brilliant being the first one; 
knots correspond to the oblique shocks in the pressure maps. 
Increasing the inclination angle, the dominant contribution to emission is due to relatively slow 
material, whose fraction in the first part of the jet is very low, so that emission 
is almost absent, apart from the first knot that is
still visible at lower brightness. The fraction of slow material starts to increase at 
a distance of about $15$ radii where the jet appears 
to have a sudden increase in opening angle, due to the formation of 
a thick layer of slowly moving material which dominates the emission at larger
distances. We can also observe some  limb brightening, explained by the
emission coming from the slow layer surrounding the deboosted
relativistic core. 

\begin{figure} 
\includegraphics[width=\hsize]{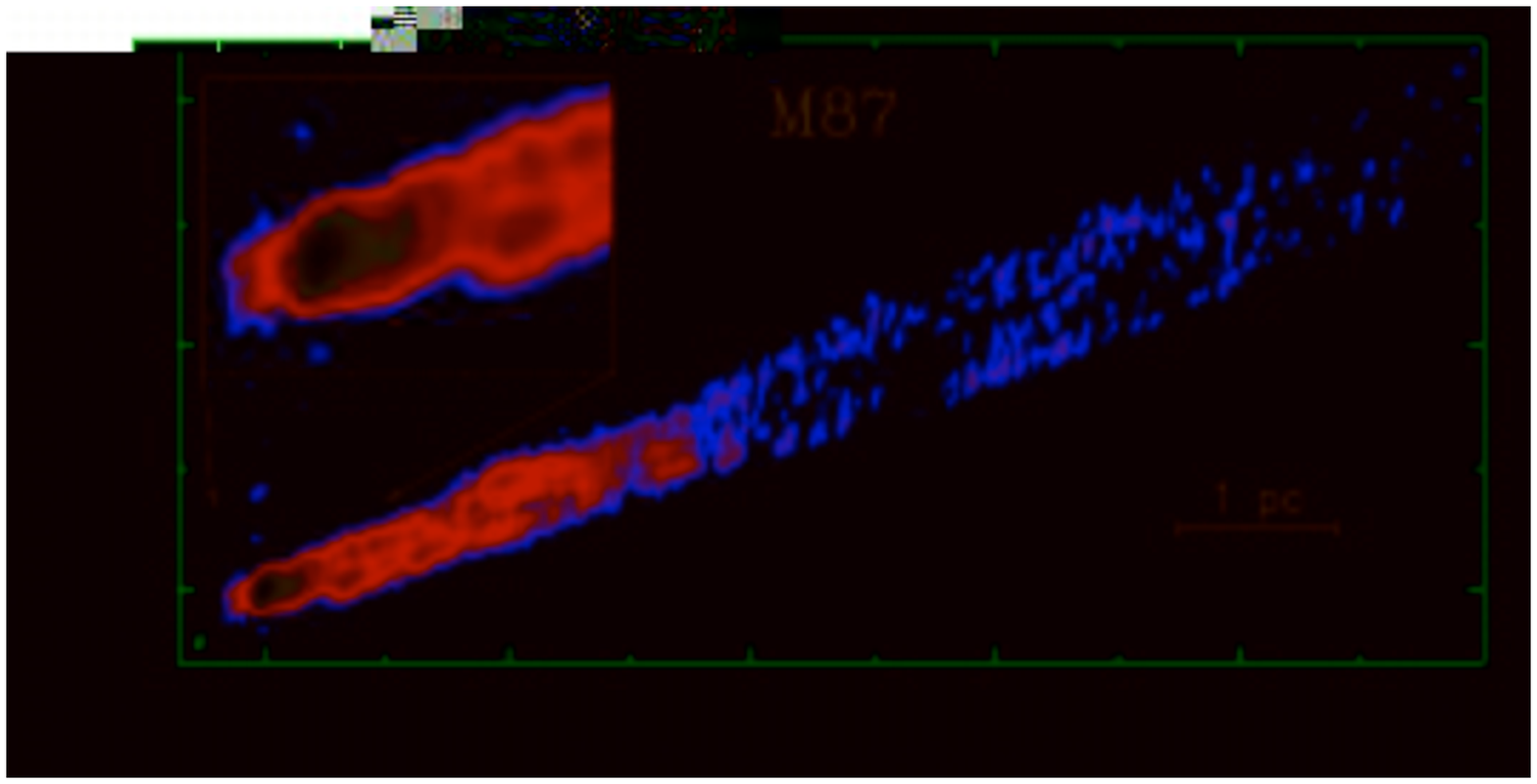} 
\caption{\footnotesize VLBA radio image at 2 cm of  M87 \citep{Kovalev07}.}
\label{fig:radio1}
\end{figure}

\begin{figure} 
\includegraphics[width=\hsize]{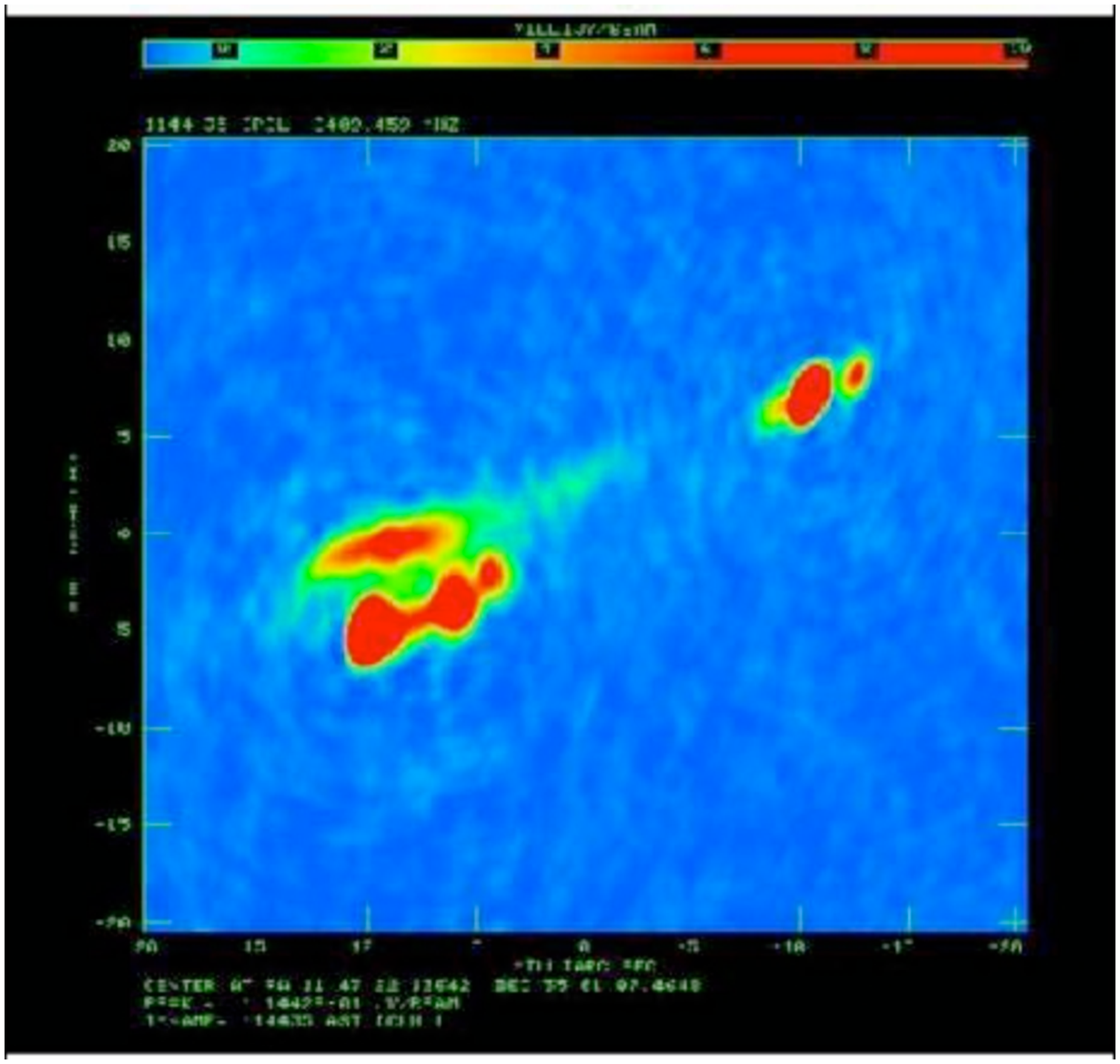} 
\caption{\footnotesize VLBI image at 8.4 GHz of B2 1144+35 \citep{Giovannini07}.}
\label{fig:radio2}
\end{figure}

We can compare these two synthetic maps with the
radio maps of two typical FR-I sources
shown in Figs. \ref{fig:radio1} and \ref{fig:radio2}. The VLBA map of M87 can be 
compared with our synthetic map at low inclination angle, which agrees with 
observational estimates.
Instead the VLBI map of B2 1144+35 can be compared with our synthetic 
map at inclination of $60^\circ$, although the estimated inclination 
\citep{Giovannini07} for this source is $33^\circ \pm 7^\circ$. 
One of the possible origins of this discrepancy lies in the fact that the
Lorentz factor of the slow layer in B2 1144+35 is estimated to be around 2.9 
(the fast spine has $\gamma \sim 15$) corresponding to $\beta \sim 0.94$,
while in our simulations (as shown in the previous section) the slow 
material is subrelativistic with $\beta \sim 0.3 -0.4$ thus becoming visible 
at larger angles. 

\section{Discussion and summary}
\label{sec:disc}
%
%
%
%
%
%
   
In this paper we have presented the 3D nonlinear dynamical evolution of
relativistic light jets, as a result of a perturbation introduced at the jet
inlet. The perturbation grows because of KHI and gives rise to a 
strong interaction of the jet
with the external medium with a consequent mixing and
deceleration. The two main parameters controlling the jet dynamics are
the Mach number $M$ and the density ratio $\eta$ between the ambient
medium and the jet. The Lorentz factor has been set equal to $10$ in 
all computations.

We have explored the parameter plane ($M, \eta$),
finding that the deceleration becomes more efficient
increasing $\eta$. 
A preliminary analysis of the parameter space suggests that only jets
with a large density ratio ($\eta > 10^{2}$) can undergo appreciable
deceleration, while the Mach number does not seem to play a fundamental
role in this respect. 

We have focused our attention on three extreme cases 
in the parameter plane, namely: case A with $M=3$ and $\eta=10^2$, B with 
$M = 3$ and $\eta = 10^4$ and E with $M = 30$ and $\eta = 10^2$.
The comparison of these simulations show in fact that case B (and D) undergoes
the strongest deceleration. Cases A  and E retain a high Lorentz factor spine with propagation 
velocities essentially unchanged from the injection value, while
some deceleration has been observed only in the outer layers with the 
formation of a wide transverse velocity structure. 
In case B  we observe the formation of a similar pattern although
on a much shorter distance and with a significant stronger decrease 
of the maximum Lorentz factor. 


The fact that larger values of $\eta$ (i.e. lower jet densities) 
lead to prominent deceleration may have direct astrophysical 
implications.
Observational data seems to indicate that the jet kinetic
power associated with FR-I radio sources is, on the average, $\sim 10^3$ 
times lower than in FR-II radio sources \citep{Celotti03}. On the other 
hand, there seems to be no difference in the value of the initial 
Lorentz factor in the two classes \citep{Giovannini01, CG08}. 
Since lighter jet beams imply reduced jet kinetic powers, 
our model leaves the density contrast as the most
likely candidate to account for the discrepancies in the deceleration 
process efficiency.  Using some astrophysically relevant units we
  can rescale our models and come up with a rough estimate for the
  critical value of the jet power $P^{*}_{j}$ that separates FRI from FRII
  radiosources:
  \begin{equation}
  P^{*}_{j}  \sim 1.3 \times  10^{44} \left( \frac{r_{j}}{1 pc} \right)^{2} 
  \left( \frac{\gamma_{b}}{10} \right)^{2}
  \left( \frac{n}{1 cm^{-3}} \right)
  \left( \frac{\eta^{*}}{10^{3}} \right)^{-1} {\rm erg s^{-1}}
  \end{equation}
where we assumed a jet radius of $1 pc$, an external density of $1
cm^{-3}$ which is typical at distances below $100pc$ \citep{BBC08}, 
and a critical density ratio separating the two behaviors of
order $10^{3}$. The value of $P_{j}^{*}$ is affected by many
uncertainities but is in agreement, for example, with the estimates 
given by \citet{CG01} based on the results by \citet{W99}, who gave
for $P_{j}^{*}$ the value
\begin{equation}
  P_{j}^{*} \sim 10^{44} \left( \frac{M_{BH}}{10^{8} M_{\sun}} \right)
  {\rm erg s^{-1} }
\end{equation}
where $M_{BH}$ is the mass of the central black hole.

Considering the case with $\eta = 10^{4}$  we obtained a
spine-layer structure similar to that deduced from observations. 
Looking at the synthetic maps produced from the simulations, it is evident 
that many of the salient features are fairly well reproduced. 
The main difference lies in the terminal  
Lorentz factor of the slow layer, typically smaller than the 
observational estimates.
In view of this first promising result we intend to further pursue
this investigation, trying to better constrain jet parameters and
introducing another essential ingredient: the magnetic field.

\begin{acknowledgements}
The numerical calculations
have been performed at CINECA in Bologna, Italy, thanks to 
INAF. This work has been partly supported by MIUR and 
the ASC FLASH Center at the University of Chicago.
\end{acknowledgements}

\end{document}